\documentclass[12pt]{article}

\pdfoutput=1

\usepackage{epsfig,latexsym,amsfonts,amsmath,amsthm,amssymb,amsbsy,multirow,slashed,wasysym,textcomp,subfigure,hyperref,wrapfig}
\usepackage[font={footnotesize,sl},bf]{caption}

\def\be{\begin{equation}}
\def\ee{\end{equation}}
\def\bea{\begin{eqnarray}}
\def\eea{\end{eqnarray}}

\newcommand{\eal}[1]{\be \begin{aligned} #1 \end{aligned}\ee} 
\newcommand{\eqn}[1]{\be #1 \ee}

\newcommand{\beq}{\begin{eqnarray}}
\newcommand{\eeq}{\end{eqnarray}}

\usepackage{color}

\long\def\new#1\endnew{{\bf #1}}		\long\def\del#1\enddel{}

\def\calh{{\cal{H}}}

\def\kt{\tilde{k}}

\newcommand{\bQ}{ Q}
\newcommand{\pQ}{q}

\newcommand{\f}[2]{\frac{#1}{#2}}
\newcommand{\nn}{\nonumber\\}

\numberwithin{equation}{section}

\begin{document}
 
\begin{titlepage}

\begin{flushright}
IPhT-T12/062\\
\end{flushright}
\bigskip
\bigskip

\centering{\Large \bf Non-extremal Black Hole Microstates:}
 \\ \medskip
 \centering{\Large \bf Fuzzballs of Fire or Fuzzballs of Fuzz ? }\\
\bigskip
\bigskip
\bigskip
\centerline{{\bf Iosif Bena, Andrea Puhm, Bert Vercnocke}}
\bigskip
\centering{Institut de Physique Th\'eorique, \\ CEA Saclay, 91191 Gif sur Yvette, France\vspace{0.4cm}}
\bigskip
\bigskip
\centerline{{ iosif.bena@cea.fr,~andrea.puhm@cea.fr,~bert.vercnocke@cea.fr} }
\bigskip
\bigskip

\begin{abstract}

  We construct the first family of microstate geometries of
  near-extremal black holes, by placing metastable supertubes inside
  certain scaling supersymmetric smooth microstate geometries. These
  fuzzballs differ from the classical black hole solution
  macroscopically at the horizon scale, and for certain probes the fluctuations between various fuzzballs will be visible as thermal noise far away from the horizon. We discuss whether these fuzzballs appear to infalling observers as fuzzballs of fuzz or as fuzzballs of fire. The existence of these solutions suggests that the singularity of   non-extremal black holes is resolved  all the way to the outer horizon
  and this ``backwards in time'' singularity resolution can shed light on the resolution of spacelike cosmological singularities.

\end{abstract}

\end{titlepage}

\setcounter{tocdepth}{2}
\tableofcontents

\section{Introduction}

According to the fuzzball proposal \cite{Mathur:2005zp, Bena:2007kg,
  Mathur:2008nj, Balasubramanian:2008da, Skenderis:2008qn,
  Chowdhury:2010ct}, black holes are a coarse grained-description of
an ensemble of horizonless microstate configurations that have the same
mass, charges and angular momenta as the classical black hole, but differ from it at the scale of the horizon.
One has by now succeeded to construct very large classes of microstate geometries \cite{Bena:2005va,Berglund:2005vb,Bena:2006kb,Balasubramanian:2006gi,Bena:2007qc, Bena:2010gg,Niehoff:2012wu} for supersymmetric black holes, as well as for
non-supersymmetric extremal black holes
\cite{Goldstein:2008fq,Bena:2009en,Bena:2009fi, Bena:2009en, Dall'Agata:2010dy,
  Bossard:2011kz,Vasilakis:2011ki}, and these geometries can be thought of as describing
the various channels for the resolution of the timelike singularity
inside the horizon of these extremal black holes. These resolution
channels modify the singular geometry to a {\em large} distance away
from the singularity, exactly as it happens in other well-understood
string-theoretic resolutions of timelike singularities, like
Polchinski-Strassler \cite{Polchinski:2000uf}, Klebanov-Strassler \cite{Klebanov:2000hb} or LLM
\cite{Lin:2004nb,Bena:2004jw}. This picture is also supported by analyzing
the physics of instabilities 
\cite{Penrose:1968ar, Brady:1995ni, Dafermos:2003wr, Poisson:1990eh,Marolf:2010nd} 
inside the horizon of extremal black holes.

On the other hand, the scale of the resolution of the singularity of
non-extremal black holes is much harder to estimate. The fuzzball
proposal and the yearning to solve the black hole information paradox
(see \cite{Mathur:2011uj} for recent work) would have the black hole
singularity resolved all the way to the outer horizon, {\em backwards
  in time} from the singularity. The recent ``firewall'' arguments of
\cite{Almheiri:2012rt,Susskind:2012ey} appear to lead in the same direction\footnote{For other related works see \cite{Bousso:2012as,Nomura:2012sw,Harlow:2012me,Mathur:2012jk,Chowdhury:2012tr}.}.

\begin{figure}[ht!]{
\subfigure[Extremal black holes.]{
\hspace{-0.2cm}
 \includegraphics[width=0.43\textwidth]{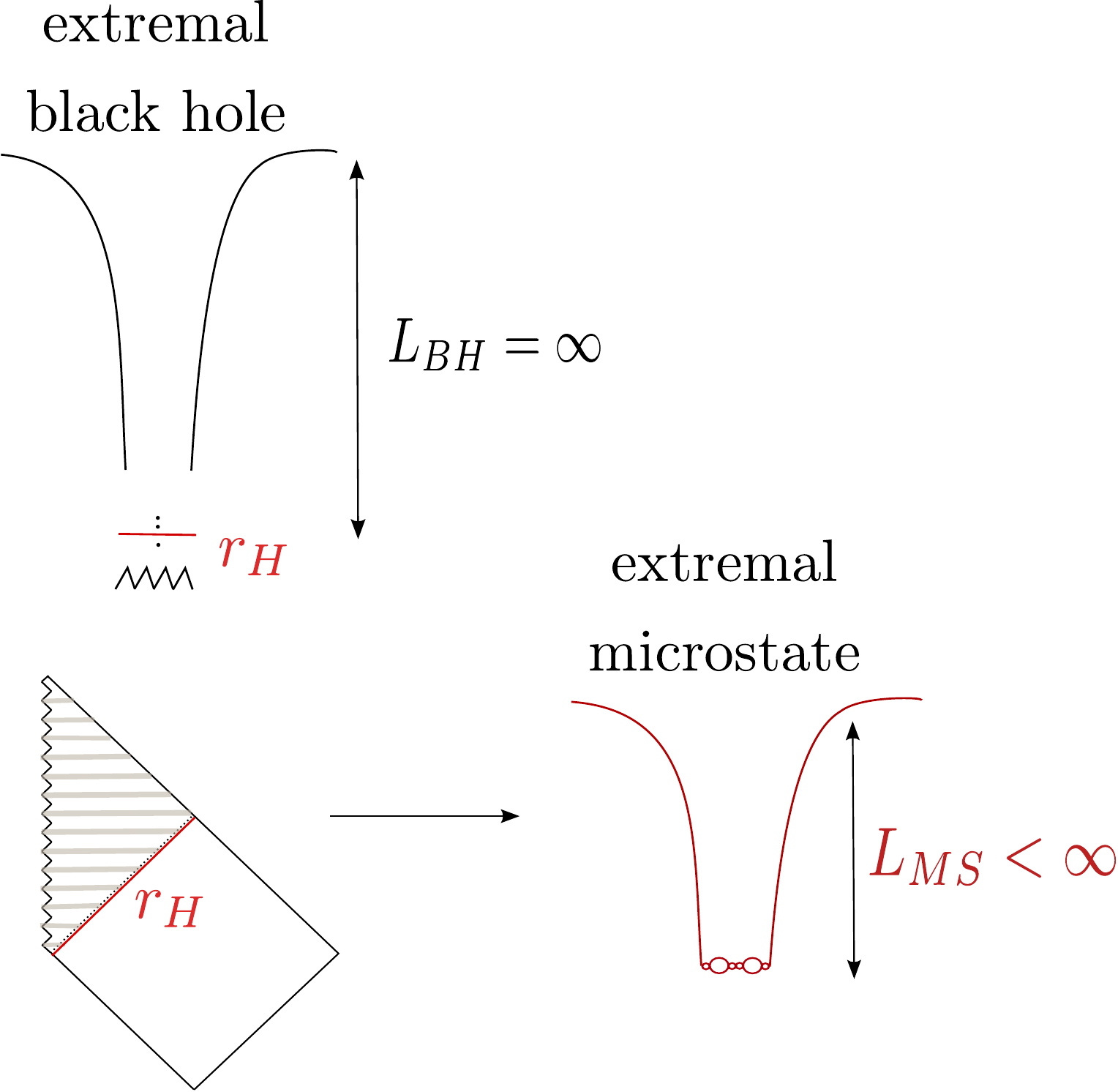}
} \hspace{0.8cm}
\subfigure[Non-extremal black holes.]{
 \includegraphics[width=.5\textwidth]{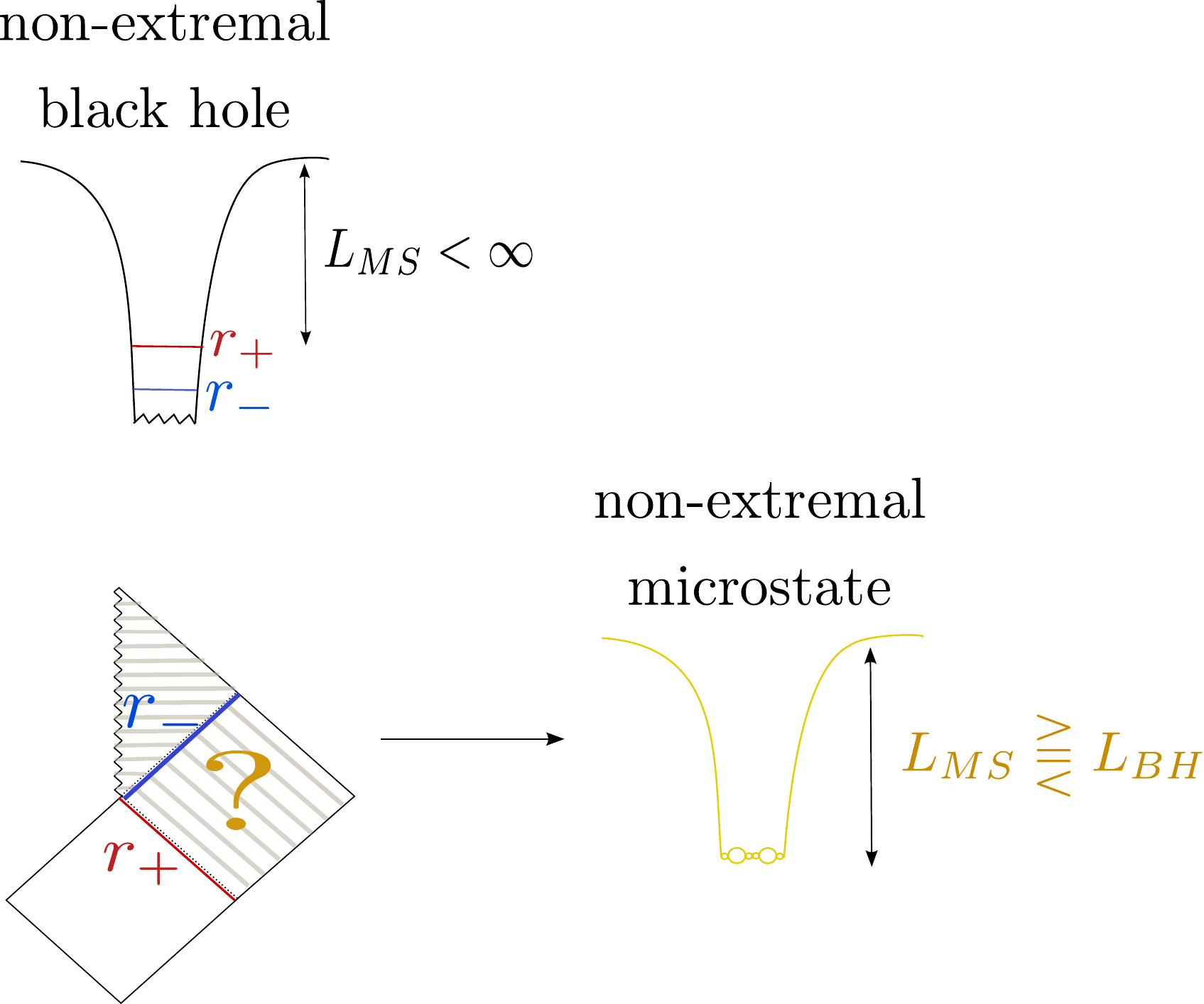}
}
\caption{\small{Singularity resolution scale.}}
\label{fig:singresol}}
\end{figure}

However, if one is to simply extrapolate the
extended evidence for extremal black hole fuzzballs to non-extremal
ones, it is well-possible that the
timelike singularity of non-extremal black holes is only resolved to
the scale of the inner horizon, and that the region between the inner
and the outer horizon is still described by the classical black hole
solution. This second possibility would not solve the information
paradox, but since it does not involves backwards-in-time singularity
resolutions it is much easier to the palate than the fuzzball/firewall
proposals. An illustration of the two possibilities is given in
figure \ref{fig:singresol}.

To address the question at which scale the singularity resolution
happens, one needs to attack the formidable task of constructing
non-extremal black hole microstate geometries, which is highly
nontrivial. Only two solutions are known: JMaRT
\cite{Jejjala:2005yu,Giusto:2007tt,AlAlawi:2009qe} and the running-Bolt
\cite{Bena:2009qv, Bobev:2009kn}; they are very non-generic, and
their generalization is nowhere in sight. In \cite{Bena:2011fc} we
argued for a way to bypass these limitations, and construct instead
microstates of near-extremal black holes by adding probes to extremal
BPS geometries. We have found that supertubes placed in generic
bubbling solutions can have metastable vacua, that can decay into the
supersymmetric ones by brane-flux annihilation, exactly as it happens
when one places antibranes \cite{Kachru:2002gs} in the Klebanov-Strassler
geometry \cite{Klebanov:2000hb}.

In this paper we want to take this technology one step further, and to
use metastable supertubes to systematically construct microstates of
near-extremal black holes. We start from supersymmetric microstate
geometries that have the same mass, charges and angular momentum as a
supersymmetric three-charge black hole, and have a very long throat
(hence they correspond to a scaling solution from the perspective of
4D supergravity \cite{Bena:2006kb,Bena:2007qc,Bates:2003vx}). We construct
near-extremal black hole microstate solutions by placing metastable
supertubes in these supersymmetric microstate geometries.

There are two ways to obtain such long-throat supersymmetric
solutions. The first is to consider a general scaling multicenter solution
and tune the length of the throat by moving the centers near each
other \cite{Denef:2002ru,Bates:2003vx,Bena:2007qc}. The other is to keep the centers aligned on an
axis, and to bring them closer and closer by tuning their charges by
hand \cite{Bena:2006kb}\footnote{This method has also been used to obtain extremal non-supersymmetric scaling solutions \cite{Vasilakis:2011ki}.}. The advantage of the second
approach is that it produces five-dimensional solutions 
with $U(1) \times U(1)$ invariance, and
in these solutions the physics of metastable supertubes is under much
better control than in scaling solutions with less symmetry.

As we discussed in \cite{Bena:2011fc}, the supersymmetry of solutions
with metastable supertubes is broken by the relative orientation of the
electric charges of the supertube with respect to the
solution. Furthermore, we will consider supertubes whose charges are
much smaller than those of the background, so we expect generically
that their backreaction will give smooth solutions with long throats,
that have more mass than charge, and hence are microstates of
\emph{non-extremal} black holes. 

Indeed, it was shown in
\cite{Bena:2008dw} that in the 6D duality frame where the
supertube charges correspond to D1 and D5 branes, a supertube in a
bubbling solution backreacts into a smooth supergravity solution. 
The smoothness is ensured by certain conditions near the supertube, which
are identical to those coming from minimizing the DBI probe supertube
action; thus we expect that the backreaction of a probe supertube at its minimum,
supersymmetric or not, will always give a smooth solution. Another way
of seeing this is to recall that the supertube is an object that
locally preserves 16 supersymmetries, and all these objects can be
dualized into fluxed D6 branes, whose eleven-dimensional uplift is
smooth \cite{Balasubramanian:2006gi}; for metastable supertubes all
these supersymmetries are incompatible with those of the background,
but this is not something that is visible in the near-supertube
region, and hence does not affect the smoothness. 

One can worry that the extra supertube charges, though small, may
disturb the delicate balance of charges needed to create a long
throat. Indeed, in a long throat the leading contributions to the
bubble equations cancel, and the supertube contributions may end up
being of the same order or larger than the subleading leftovers. However,
this is not a problem; even if a supertube changes significantly the length of a
throat, one can always tune the flux between cycles by a tiny amount
to change this length back to the original one, and this gives a very
small correction to the overall charges of the solution. 

Having obtained large classes of microstates of near-extremal black
holes, we can go on and attack the much harder problem of figuring out
what is the physics of these microstates. Indeed, according to the
fuzzball proposal one expects the microstates to have the same size as
the black hole, but there are various scenarios of how this can
happen. It is possible that microstates extend only microscopically
away from the horizon, and hence, as suggested in \cite{Almheiri:2012rt}, they could
give nothing more than a realization of a stretched
horizon. Alternatively, the fuzzballs can differ from the black hole
on a scale comparable to that of the horizon, and hence give very
different physics.

These questions do not make sense for BPS and extremal black holes,
and cannot be answered using the BPS black hole microstates constructed
so far. Indeed, the thickness of BPS throats is completely determined
by the charges, and hence a fuzzball and a black hole that have the
same charges automatically have throats of equal thickness.
Furthermore, the length of the throat of the black hole is infinite,
while the length of the throat of the fuzzballs is always very large
but finite\footnote{The only way to figure out whether a long BPS microstate
is typical is to compare its length to the mass gap of the typical
microstate in the dual CFT, and this comparison indicates that long
microstates whose angular momentum is of order one belong indeed the
sector where the typical microstates live
\cite{Bena:2006kb,Bena:2007qc,deBoer:2008zn}.}, which does
not allow for a meaningful size comparison. On the other hand,
near-extremal black holes have throats of {\it finite} length, which
one can compare with the throat lengths of the family of fuzzballs we
construct. The thicknesses of the throats are still automatically
equal, because they are mostly controlled by the charges in the
near-extremal limit.

As we will see, we are able to construct microstates
whose throat has the same length as that of a non-extremal black hole, but we can also obtain with equal ease 
microstate geometries that have the same mass and
charges as the non-extremal black hole, but whose throats are longer or
shorter. As far as our construction is concerned, there appears to
be no dynamical reason why throats of the same length as the black hole
are preferred over longer or shorter ones, and this indicates that the
fuzzballs of non-extremal black holes will not differ from the black
hole only at microscopic distances from the horizon\footnote{Most
  likely there will be a distribution of fuzzballs of various lengths,
  and the length of the typical ones will come out to be same as the
  length of a black hole by some entropy enhancement reason
  \cite{Bena:2008nh}.}. 

The fuzzball geometries we construct can be used to extract other
pieces of physics that have been inaccessible until now. For example,
one can use a KKLMMT-type argument \cite{Kachru:2003sx} to find the forces with which our
fuzzballs attract various D-branes, and compare these forces to those
of the corresponding black hole. One can also compute the tunneling probabilities of
the metastable supertube to the supersymmetric minimum, and compare
this to the Hawking radiation rate of the near-extremal black hole; we leave this for future work.

This paper is organized as follows. In section \ref{ch:supertube} we
recall the Hamiltonian of supertubes in three-charge backgrounds. We then focus on a
scaling solution with seven centers in section \ref{ch:example} and
plot the potential of a typical probe supertube in this background. In
section \ref{ch:physics} we discuss the interpretation of our configurations as microstates (or fuzzballs) of non-extremal black holes, and compare their properties to those of black holes. In section \ref{ch:conclusions} we discuss whether the fuzzballs we construct appear as fuzzballs of fuzz or as fuzzballs of fire to incoming observers, we speculate on the implication of this work for our understanding of the resolution of spacelike singularities in String Theory, and discuss possible future directions. In appendix \ref{app:3chargescaling} we review the construction and the physics of smooth BPS black hole microstate solutions, in appendix  \ref{app:BlackHole}, we recall the geometry of the non-extremal black hole, and in appendix  \ref{app:Depth} we present an approximation that allows us to compare very easily the lengths of a black-hole throat and of a fuzzball throat. 


\section{Supertubes in scaling backgrounds}
\label{ch:supertube}

In this section, we review the potential for supertubes in a
supersymmetric three-charge background. We focus on deep supersymmetric microstate geometries (smooth, horizonless three-charge solutions with scaling behavior) and explain how the scaling effects the supertube potential.

\begin{figure}[ht!]{ \centering{
\includegraphics[width=0.6\textwidth]{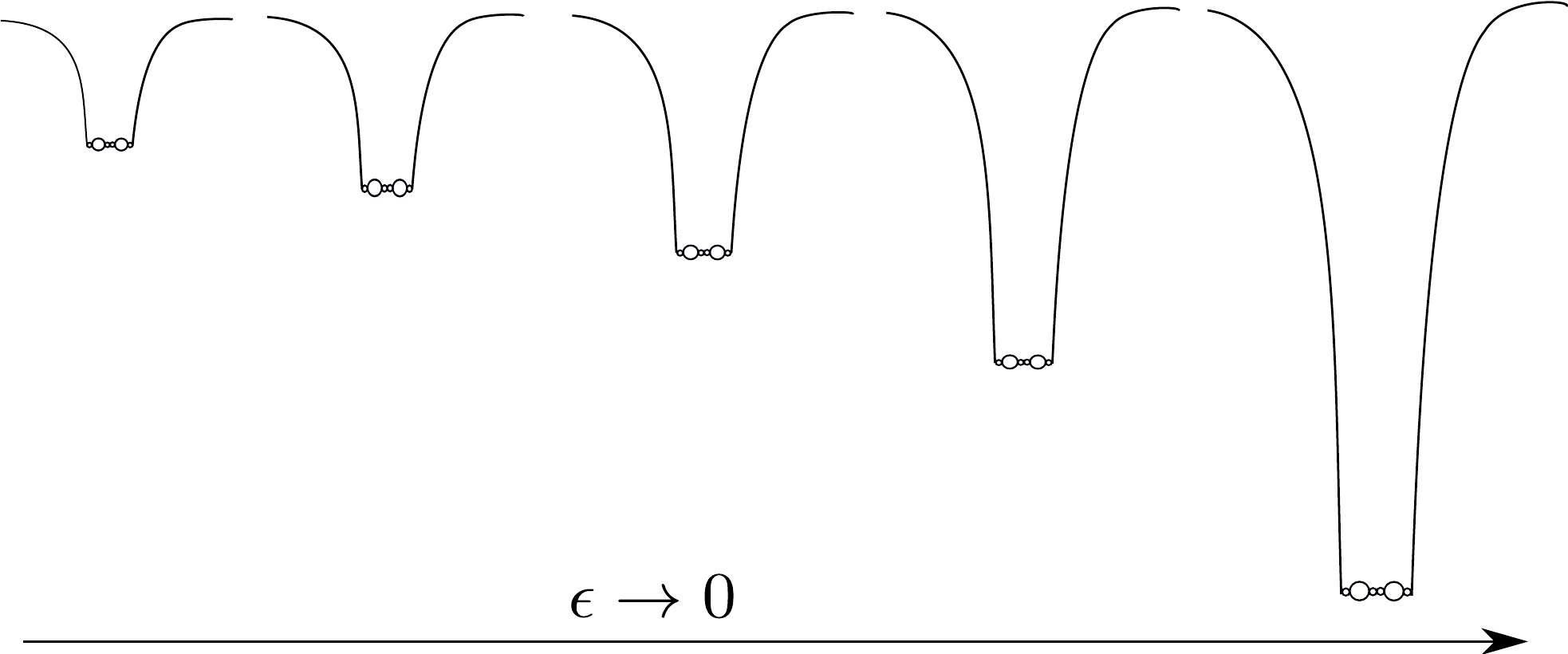}

\caption{Heuristic picture of scaling microstate geometries.}}}
\end{figure}

\subsection{Supertubes in three-charge backgrounds}

Consider a supersymmetric background geometry with three charges and three dipole charges, of the type that describes black holes, black rings and their microstate geometries. The metric in the M-theory duality frame in which the three charges correspond to M2 branes wrapping orthogonal $T^2$'s inside $T^6$ is \cite{Bena:2004de,Gutowski:2004yv}:
\bea
ds_{11}^2 &=& -(Z_1 Z_2 Z_3)^{-2/3}(dt + k)^2 + (Z_1 Z_2 Z_3)^{1/3} ds_4^2 
+(Z_1 Z_2 Z_3)^{1/3}\sum_{I=1}^3 \frac{ds_I^2}{Z_I}
\label{11dgeometry}\\
F_4 &=& \sum_{I=1}^3 dA^{(I)}\wedge \omega_I\,,\qquad dA^{(I)} = -d[Z_I^{-1} (dt+k)]+\Theta^{(I)}\,.
\eea
where $ds_I^2$ and $\omega_I$ are unit metrics and volume forms on the three orthogonal $T^2$'s and $ds_4^2 $ is the metric of a hyper-K\"{a}hler base space. Supersymmetry requires the two-forms $\Theta^{(I)}$ to be self-dual on the base. When the hyper-K\"ahler space is Gibbons-Hawking (GH) or Taub-NUT:
\bea
d s^2_4 &=& V^{-1} (d \psi + A)^2 + V ds_3^2\qquad \text{with} \qquad d A = \star_3 d V,
\label{eq:GH}
\eea
where $ds_3^2$ is the flat metric on $\mathbb{R}^3$, the solution is completely determined by specifying 8 harmonic functions $V,K^I,L_I,M$ in the GH base \cite{Gauntlett:2004qy,Elvang:2004ds}. 
The harmonic functions can have sources on an arbitrary number of positions in $\mathbb{R}^3$. The warp factors and rotation one-form are given by
\bea
Z_I &=& L_I + \tfrac 12  C_{IJK}V^{-1} K^J K^K\,,\\
k &=& \mu (d\psi + A) + \omega  \,,
\label{eq:warprot}
\eea
with $C_{IJK}=|\epsilon_{IJK}|$ and 
\bea
\mu &=&  \tfrac 16 C_{IJK}V^{-2} K^I K^J K^K +\tfrac  12 L_I K^I + M\,,\nonumber\\
\vec \nabla \times \vec \omega &=& V \vec \nabla M - M \vec \nabla V + \tfrac 12 (K^I \vec \nabla L_I - L_I \vec \nabla K^I)\,.
\eea
Note that the inverse of the warp factors $Z_I$ are also the electric potentials for the four-form and hence they determine the M2 charges at each background center.

In \cite{Bena:2011fc} we found the Hamiltonian of a two-charge supertube in such a multicenter three-charge background with a Gibbons-Hawking base. The two charges $\pQ_1$ and $\pQ_2$ of the supertube are parallel to those of the background and correspond to M2 branes along the first and second $T^2$. The dipole charge, $d_3$, corresponds to an M5 brane extended along those two tori wrapping the fiber of the Gibbons-Hawking space. The Hamiltonian is:
\eal{
\calh &= \frac{\sqrt{Z_1Z_2Z_3/V}}{d_3R^2}\sqrt{\Big(\tilde \pQ_1^2 +d_3^2 \frac{R^2}{Z_2^2}\Big)\Big(\tilde \pQ_2^2 + d_3^2\frac{R^2}{Z_1^2}\Big)}
+ \frac{\mu V^2}{d_3R^2}\tilde \pQ_1 \tilde \pQ_2 - \frac 1 {Z_1} \tilde \pQ_1 - \frac 1 {Z_2} \tilde  \pQ_2 -  \frac{ d_3 \mu}{Z_1Z_2} + \pQ_1 + \pQ_2 \,,\label{eq:Hamiltonian}
}
where we have introduced 
\be
\tilde \pQ_1 \equiv  \pQ_1 + d_3  (K^2/V - \mu/Z_2)~,~~~\tilde \pQ_2 \equiv  \pQ_2 + d_3  (K^1/V - \mu/Z_1)\,,
\ee
and $R$ is proportional to the size of the Gibbons-Hawking fiber
\be
R^2 \equiv Z_1Z_2Z_3/V - \mu^2\,.
\ee
The harmonic functions $K^1$ and $K^2$ encode two of the three dipole
moments of the background. The minima of the potential
determine the position on the GH base of (meta)stable supertubes in a
given three-charge background. Depending on the relative orientation
of the M2 charges of supertube and the background, the minima of the
potential will be supersymmetric (with energy $V_{BPS}=\pQ_1+\pQ_2$)
or non-supersymmetric.

\subsubsection{Supertubes in scaling backgrounds}

A scaling background is a bubbling configuration that has a set of GH
points that can approach each other arbitrarily close. As the points
get closer together, the solution develops an ever deeper throat and
looks more and more like the black hole with the same asymptotic
charges. See appendix \ref{app:3chargescaling}. Deep scaling solutions
are dual to states that belong to the same CFT sector as the typical
microstates, that give the leading  contribution to the black hole
entropy \cite{Bena:2006kb,Bena:2007qc,deBoer:2008zn}.

For a given set of charges, a scaling limit exists if we can find a solution to the bubble equations \eqref{eq:bubble_quantized} or \eqref{eq:bubbleeq} for
\eqn{ 
r_{ij} =  \ \epsilon \,\tilde r_{ij} \qquad \text{with} \quad \epsilon \rightarrow 0.\label{eq:ScalingLimit}
}
When such a solution can be found, all distances in the GH base scale to zero, but the physical size of the bubbles and ratios between distances are preserved throughout the scaling $\epsilon \to 0$,  because the warp factors along the bubbles diverge appropriately.

When we focus on scaling backgrounds we find that the Hamiltonian has a similar scaling.  By rescaling the coordinates on the 3d base as $\vec r = \epsilon\, \vec {\tilde r}$, and taking the limit $\epsilon \to 0$, the Hamiltonian scales as
\eqn{ 
\calh\big{(}\epsilon\, \vec {\tilde r}\,\big{)} = \epsilon \, \calh\big{(}\vec{\tilde r}\,\big{)} + \mathcal{O}(\epsilon^2)\label{eq:ScalingHamiltonian}
}

As mentioned in the Introduction, there exist two ways of obtaining a
scaling solution. The first is to consider a set of $N$ centers whose
charges allow for scaling behavior; they satisfy $N-1$ bubble
equations, and their $2N-2$ dimensional moduli has a region where the
points come together, and the fully backreacted solution develops a
long throat \cite{Bates:2003vx,Bena:2007qc}. The second way is to insist that
the centers be collinear -- their positions are now parameterized by $N-1$
variables that are completely determined by the $N-1$ bubble
equations -- and force the centers to scale by tuning by hand some of
the flux parameters on the centers or some of the moduli of the
solution \cite{Bena:2006kb}. 

We will use the second approach, essentially because it gives much
more control on the dynamics of the supertube. If one adds a supertube to a
scaling solution whose centers are not collinear, the energy of the
supertube depends on the length of the throat, and can change as the
centers move in the moduli space. On the other hand, in a  $U(1)
\times U(1)$ invariant solution the centers are collinear and hence
frozen, and if the supertube charges are smaller than those of the
other centers, the physics of the metastable supertube is expected to
be captured by its probe action in the background. 

Thus, in the examples in the next section, we focus on scaling
solutions where all the GH points are collinear and we will `turn the
knob' of the scaling control parameter $\epsilon$ by tuning one of the
charges $k_i^I$.

\section{A seven-center scaling solution}
\label{ch:example}

\begin{figure}[ht!]
\begin{center}
\includegraphics[width=.7\textwidth]{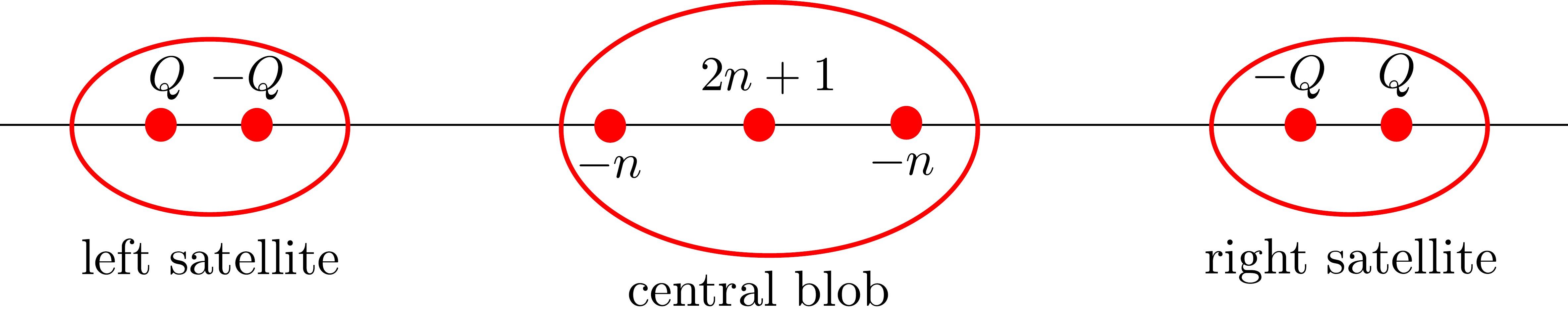}
\end{center}
\caption{Our setup consists of a central blob of three centers and two satellite blobs of two centers each, with the GH charges as given in the figure.\label{fig:PincerBlobs}}
\end{figure}

In this section, we analyze the minima the probe supertubes in a
``pincer'' supersymmetric scaling background (inspired from
\cite{Bena:2006kb}) whose centers are colinear in $\mathbb{R}^3$ and have $J_L=0$.
This pincer solution contains a central `blob' of total GH charge one,
as well as two symmetric satellite blobs of GH charge zero.  For
computational ease, we take a configuration that is made up out of a
total of seven points on the GH base: a central blob made from three
points, of GH charges $-n, 2n +1,-n$, and two satellites with two
points that have GH charges
$-Q$ and $+Q$. The configuration, depicted in Figure
\ref{fig:PincerBlobs},  is $\mathbb{Z}_2$ symmetric, and hence has
$J_L=0$ by construction.

One can then choose fluxes between the various GH centers such that
the total configuration has the charges of a BPS black hole with a
macroscopically-large horizon area. The particular choice of fluxes
that ensures that a five-point solution has no CTC's was obtained in
\cite{Bena:2006kb} by tediously analyzing blob mergers, and
our choice is simply the $\mathbb{Z}_2$ symmetrization of that choice.

\subsection{Background data}

\begin{figure}[ht!]{ \centering{
\includegraphics[width=0.6\textwidth]{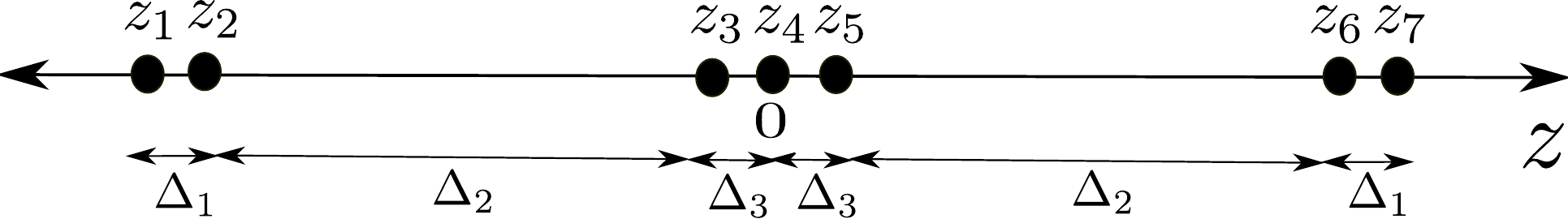}
\caption{Schematic picture of our microstate configuration.\label{fig:pincerconfig}}}}
\end{figure}

We choose a cylindrical coordinate system $(\rho,z,\theta)$ in three
dimensions, where $z$ runs along the axis through the centers and
$\rho,\theta$ are polar coordinates in the orthogonal plane.  Since we
have cylindrical symmetry, the solution only depends on the
coordinates $z$ and $\rho$. The seven centers are put on the $z$-axis
and are numbered $z_1 \ldots z_7$ as in figure
\ref{fig:pincerconfig}. We choose the GH charges to be
\eqn{ 
v_1=20, \; v_2=-20, \qquad v_3=-12, \; v_4=25, \; v_5=-12, \qquad v_6=-20, \; v_7=20.
\label{eq:fixGH}
}
The flux parameters of the central blob are chosen as
\eqn{ 
k^1_i=\frac{5}{2} |v_i|, \qquad k^2_i=\hat{k} |v_i|, \qquad k^3_i=\frac{1}{3} |v_i|\,,\qquad i=3,4,5\,,
\label{eq:fixflux}
}
and those of the satellites are
\begin{alignat}{4}
k^1_1 &=1375\,,\qquad k^2_1 &= -\frac {1835}2 + 980 \hat k\,,\qquad k_1^3&=-\frac{8360}3\,,\nonumber\\
k^1_2 &= -1325\,,\qquad k_2^2 &= \frac{1965}{2} - 980 \hat k\,,\qquad k_2^3 &=\frac{8380}{3}\,,
\end{alignat}
and their mirror image $k_7^I=k_1^I,\, k_6^I=k_2^I$. 

The charges of the harmonic functions are then a function of $\hat k$ only. For every value of $\hat k$, the bubble equations \eqref{eq:bubbleeq} fix the position of the seven centers.  One can approximate the size of the microstate by  $z_6\approx r_0$ as in \cite{Bena:2006kb}:
\be
r_0 = \frac{\widehat J_L}{8 \sum_I (k^I_6 + k^I_7)}\,,
\ee
where $\widehat J_L$ is the angular momentum contained in the centers $z_3,\ldots z_7$. In the given background, this is linear in $\hat k$ as:
\be
r_0 = \frac{56}{31}\times 10^3|\hat k - \hat k_\star|\,,\qquad k_\star \approx 3.17975\,.
\ee
We will tune $\hat{k}$ such that $r_0\to0$ and the configuration scales down into a deep throat.

In table \ref{tab:Merger} we list the relevant distances and ratios of distances for various values of the flux parameter $\hat{k}$. Starting from one set of inter-center distances the bubble equations (\ref{eq:bubbleeq}) successively determine the equilibrium distance for every value of $\hat{k}$ during the scaling. 

\begin{table}[ht!]
\begin{center}
\begin{tabular}{|c|c|c|c|c|c|c|}
\hline
Background& $\hat k$ & $z_6$&$\frac{z_6}{r_0}$& {\large $ \frac{\Delta_3}{\Delta_1}$}&{\large $\frac{\Delta_2}{\Delta_1}$}\\[1mm]
\hline
\hline
 1 & 3.08333 & 176.088 & 1.011 & 1.5464 & 6730.9 \\[1mm] \hline
 2 & 3.16667 & 23.91 & 1.01166 & 1.61449 & 6664.58 \\[1mm] \hline
 3 & 3.175 & 8.69039 & 1.01279 & 1.6215 & 6657.94 \\[1mm] \hline
 4 & 3.1775 & 4.12444 & 1.01474 & 1.62362 & 6655.95 \\[1mm] \hline
 5 & 3.178 & 3.21125 & 1.0158 & 1.62404 & 6655.55 \\[1mm] \hline
 6 & 3.17833 & 2.60246 & 1.01693 & 1.62432 & 6655.29 \\[1mm] \hline
 7 & 3.17867 & 1.99366 & 1.01874 & 1.6246 & 6655.02 \\[1mm] \hline
 8 & 3.1795 & 0.471667 & 1.04441 & 1.62531 & 6654.36 \\[1mm] \hline
 9 & 3.17967 & 0.167268 & 1.11114 & 1.62545 & 6654.22\\[1mm]
\hline
\end{tabular}
\end{center}
\caption{Distances between the points throughout the scaling process. The distances $\Delta_i$ and $z_6$ are as in Figure \protect{\ref{fig:pincerconfig}}. The parameter $\hat k$ is tuned for the scaling, all the other charges are kept fixed at their values \protect{\eqref{eq:fixGH}} and \protect{\eqref{eq:fixflux}}. It is clear that the relative distances stay approximately the same during the scaling. Also the total charges $Q_I$ and angular momentum $J_R$ stay approximately the same throughout the scaling.\label{tab:Merger}}
\end{table}

\subsubsection*{Charges and angular momenta}

The values of the electric charges and the right-moving angular momentum as defined in (\ref{eq:elcharges}) and (\ref{eq:angmomenta}) stay approximately constant throughout the scaling
\eqn{ 
Q_1 \approx 1.476\times 10^5\,,\quad Q_2 = 1.196 \times 10^5\,,\quad Q_3 \approx 1.76 \times 10^5\,,\quad  \text{and} \quad J_R \approx 1.018 \times 10^8\,.
}

Note that $Q_2$ is independent of $\hat k$. Since the
configuration is symmetric (the charges of opposite centers are the
same), the left-moving angular momentum is exactly zero
throughout the whole merger process.\footnote{For configurations with
  only one satellite this symmetry is broken and $J_L \neq 0$. Then
  $J_L$ goes to zero as the solution gets deeper and deeper. The
  end-point of such a merger is a BMPV black hole microstate with
  $J_L=0$. Only in this deep-throat limit the microstates have the
  charges of a black hole of non-zero entropy, while our background
  has the charges of a BMPV black hole throughout the scaling.} Since
the charges and angular momenta all stay nearly constant
\textit{throughout the scaling} these microstates have the charges of a
black hole of non-zero entropy in all regimes: when they are shallow
(before the scaling), when they are very deep (in the scaling limit)
and in the whole intermediate regime.

\subsection{The supertube potential}

\begin{figure}[ht!]
\begin{center}
\includegraphics[width=0.99\textwidth]{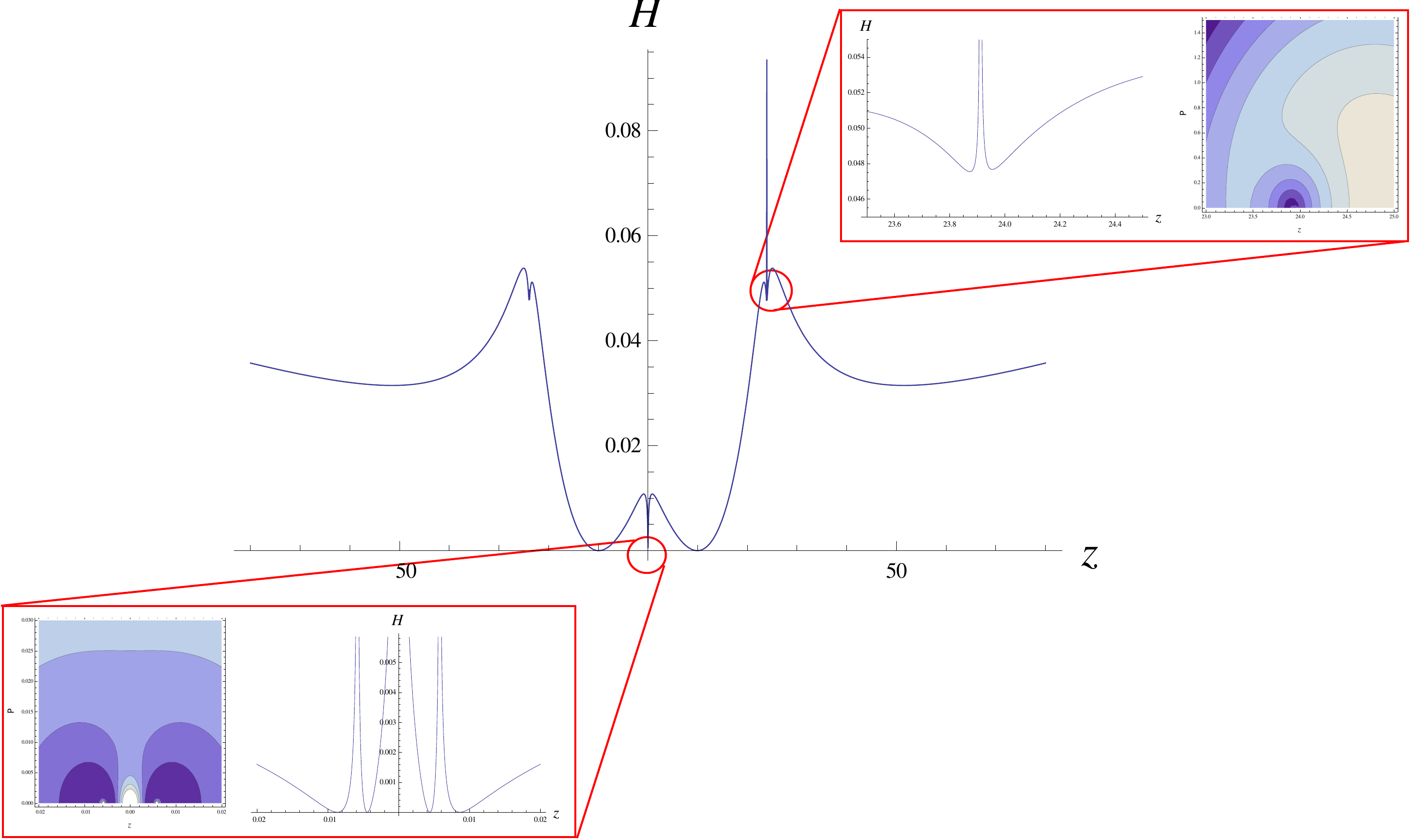} 
\end{center}
\caption{Zoom on the supertube potential for charges $(\pQ_1,\pQ_2,d_3) = (10,-50,1)$ in background 2. Note the metastable minimum near $z_6 =23.91$ (and its mirror near $z_2 =-23.91$). The contour plot shows that this minimum is of ``Mexican hat -- type'' in the $z-\rho$ plane around the center $z_6$ ($z_2$); darker colors mean lower energy. On can see that this minimum has no runaway behavior in the $\rho$ direction and hence is truly metastable. The supertube in that minimum can tunnel to a supersymmetric state. Note also that the minima near the central blob are in fact two mirror copies of a Mexican hat-type circular band of  minima, as the contour plot in the bottom left corner shows.\label{fig:Potential_1}}
\end{figure}

We plot the potential for a probe supertube in this background, with supertube charges
\be
(\pQ_1,\pQ_2,d_3) = (10,-50,1)\,.
\ee
The potential is normalized to zero for a supersymmetric minimum:
\be
\tilde \calh \equiv \calh -(\pQ_1 + \pQ_2)\,
\ee 
and we will omit the tilde in the following.
For illustrative purposes, we plot the potential of the supertube in the background `2' of Table \ref{tab:Merger} as a function of $z$ and perform a Contour plot around the minima in a plane through the $z$-axis, see Figure \ref{fig:Potential_1}. The positions of the seven centers in background 2 are
\be
z_1= -23.9136\,,\;\, z_2= -23.91\,,\;\, z_3= -0.00579078\,,\;\,  z_4= 0\,,\;\, z_5 = |z_3|\,,\;\, z_6 = |z_2|\,,\;\, z_7=|z_1|\,.
\ee
The potential has several supersymmetric minima: two lie inside the central blob, two lie just outside  and there are two more minima in between the central blob and the satellite centers at $z\approx \pm10$. There are two metastable minima close to the satellites, near $z_2$ and $z_6$. Since the setup is symmetric, we focus on the metastable minimum at $z_{\rm ms} \lesssim z_6$:
\be
z_{\rm ms} = 23.8729\,,\qquad z_6 =23.91\,.
\ee
The supertube in that minimum can tunnel to the supersymmetric state at $z \simeq 10$ via brane-flux annihilation as explained in \cite{Bena:2011fc}.
Note that the additional non-supersymmetric minima near $z= \pm 50$ as seen from Figure \ref{fig:Potential_1} are in fact saddle points and they have a runaway behavior off the axis.

In order to stay well in the probe approximation one needs to make sure that the charges of the supertube are small compared to the charges of the background (as measured by the poles of $L_I$). In particular, the metastable minimum at $z_{\rm ms}$ sits close to the centers $z_6$ and $z_7$ of the background, and we have to make sure that the charges at that position are large compared to the ones of the supertube. For the charges and flux parameters as fixed in (\ref{eq:fixGH}) and (\ref{eq:fixflux}) the background electric charges at the black ring centers are of the order $3\times 10^5$ and hence our supertube is well in the probe regime. 

\begin{figure}[ht!]
\begin{center}
\subfigure[Potential in background 2, $z_6 =23.91$]{\includegraphics[width=0.4\textwidth]{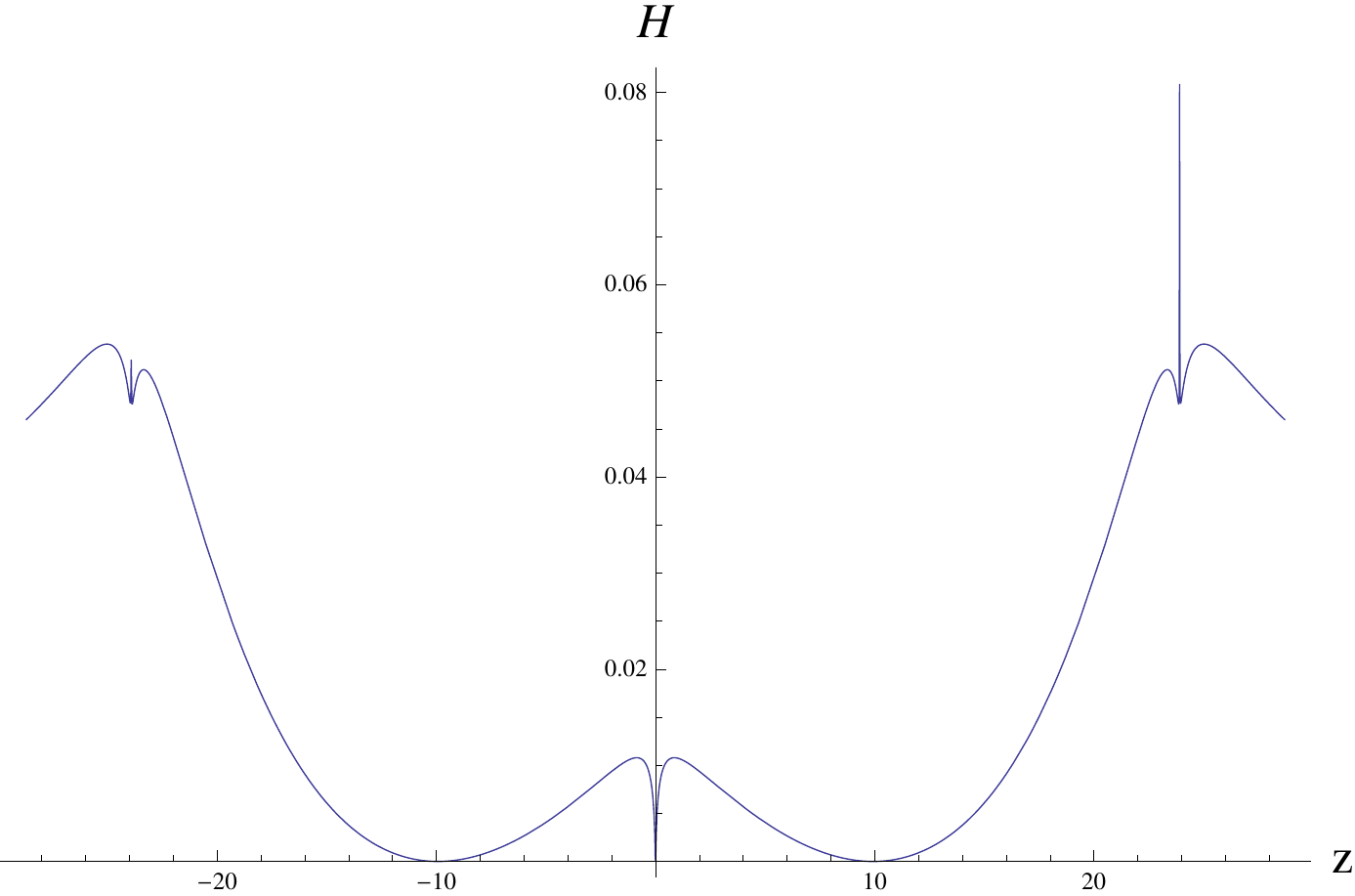}}
\hspace{.1\textwidth}
\subfigure[Potential in background 9, $z_6 = 0.167268$]{\includegraphics[width=0.4\textwidth]{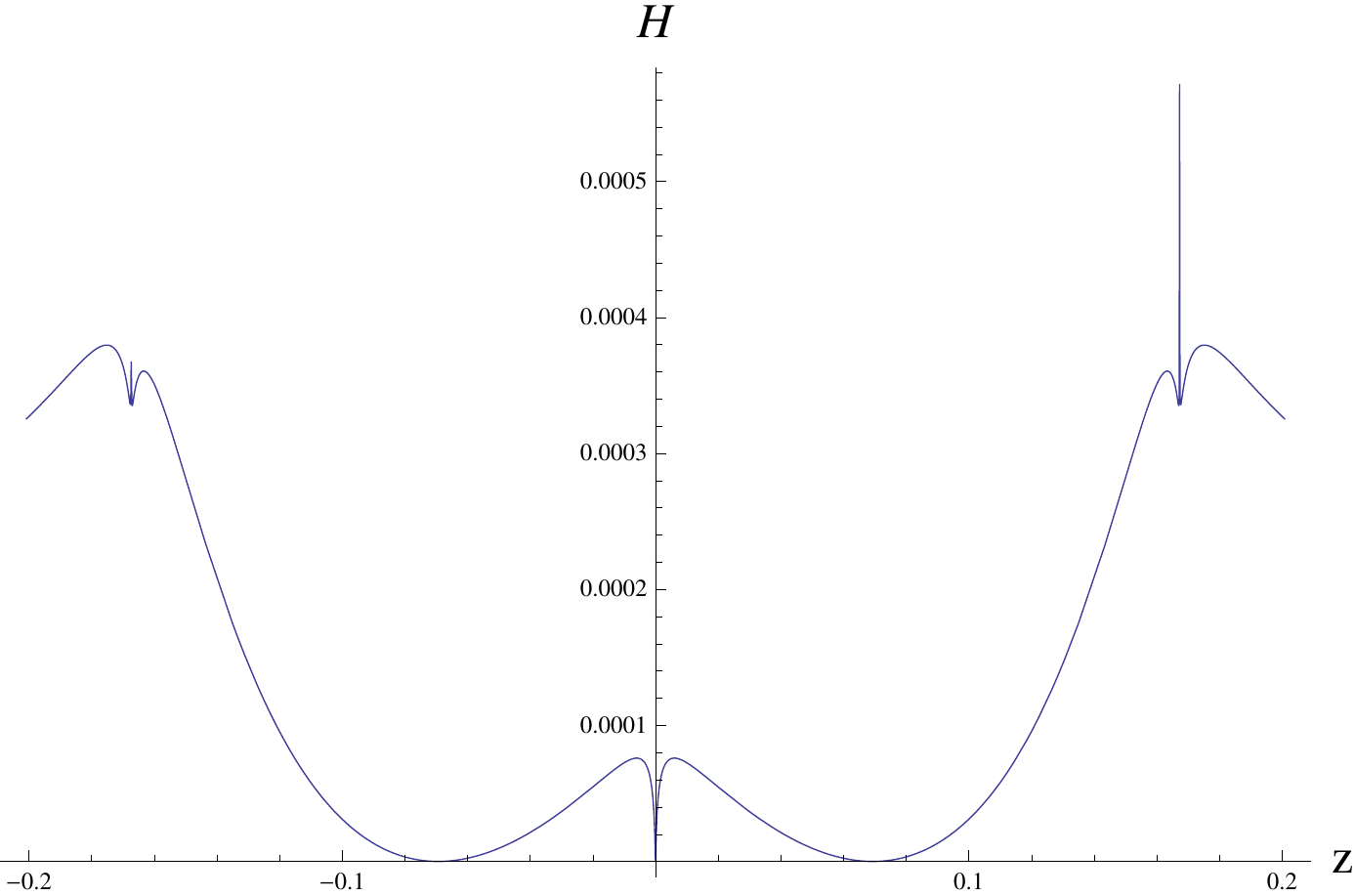}}
\end{center}
\caption{The supertube potential for charges $(\pQ_1,\pQ_2,d_3) = (10,-50,1)$ in two scaling backgrounds. The energy scales down linearly with the coordinate size  between the centers.\label{fig:Potential_2}}
\end{figure}

As explained before, the supertube potential scales down linearly with the coordinate distance between the background centers, see eq.\ \eqref{eq:ScalingHamiltonian}. As an illustration, we compare the potential for two scaling backgrounds, 2 and 9 of Table \ref{tab:Merger} in Figure \ref{fig:Potential_2}. One clearly sees the self-similarity of the potentials. Also the supertube position $z_{\rm ms}$ scales down with the throat: its relative position to the other centers stays unchanged.

\section{Non-extremal microstate throats}
\label{ch:physics}

Upon backreaction, metastable supertubes in scaling backgrounds should
become microstates of a non-extremal black hole. In this section we
want to compare the size of these microstates to the size of the
corresponding black hole, and understand the scale at which non-extremal
fuzzballs differ from the black hole.

\subsection{The idea}

One can estimate the depth of a black hole or of a fuzzball throat by integrating the radial metric
component:
\be
L = \int_{r_{\rm bottom}}^{\rm r_{neck}} \sqrt{g_{rr}} dr\,,
\ee
between the bottom and the neck of the throat. To get the depth of the
non-extremal microstate, we can evaluate this integral in the
supersymmetric background geometry since the probe supertube will not
affect the geometry too much. We then compare this to the depth of the
throat of the non-extremal black hole (a Cvetic-Youm black hole
\cite{Cvetic:1996xz}, see appendix \ref{app:BlackHole}) that has the
same charges.

\begin{figure}[ht!]
\begin{center}
\includegraphics[width=0.5\textwidth]{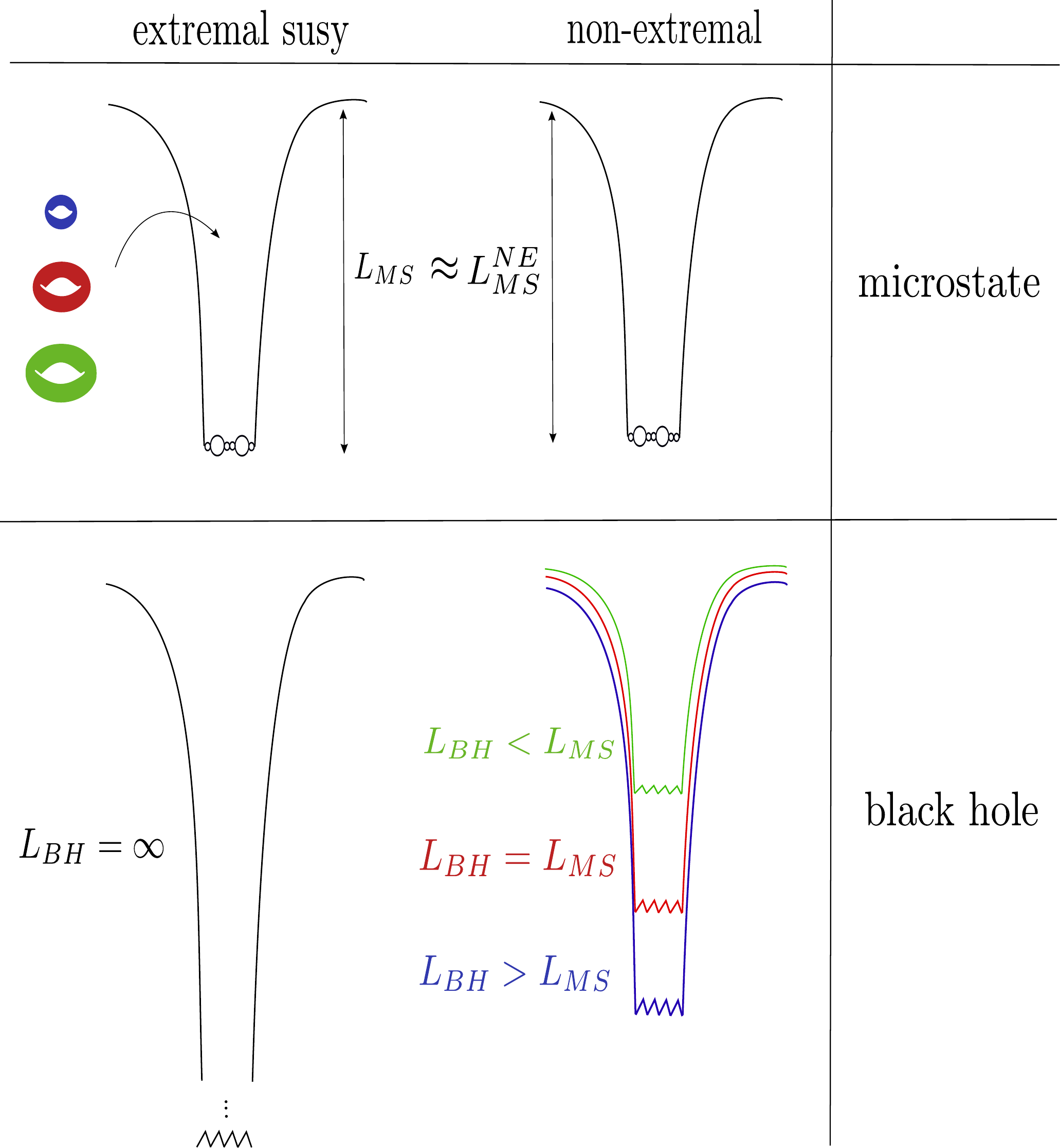}
\end{center}
\caption{The scaling of the background determines the size of the
  metastable black hole microstates. For a BPS background of a fixed
  depth, adding heavy supertubes gives a microstate of a black hole
  that has shorter throat, while adding light supertubes gives a
  microstate of a black hole with a longer throat.}
\label{fig:ThroatIdea}  
\end{figure}

The main result of this paper is that we, indeed, find microstates
that are of the same depth as the non-extremal black hole, but we also
find deeper ones and shallower ones.  This is not surprising:
Supertube probes placed in deep scaling solutions will not affect the
background geometry too much upon backreaction and the resulting
non-extremal microstate will, hence, be of the same size as the
supersymmetric background. The size of the corresponding non-extremal
black hole, however, depends on the extremality parameter which is set
by the charges of the supertube. Small supertube charges correspond to
deep black holes; increasing the tube charges takes the black hole
further away from extremality and thus makes the throat more
shallow. Hence, by tuning the supertube charges we can always find the
throat of the non-extremal black hole to be of a size comparable to
that of its microstates. This intuition is summarized in \mbox{Figure
  \ref{fig:ThroatIdea}}.

In the remainder of this section we make this intuitive picture more
precise. First, we determine the data of the non-extremal black hole
with the charges of the metastable bound states in Section
\ref{ssec:BlackHoleParemeters}. We give the depths of the black hole
and microstate throats in Section \ref{ssub:Comparing}. Since the
resulting integrals are quite complicated, we make an insightful
approximation in appendix \ref{app:Depth}.

\subsection{Non-extremal black hole parameters}\label{ssec:BlackHoleParemeters}

We begin with a supersymmetric fuzzball solution that has the charges
of a supersymmetric rotating BMPV black hole, and its mass is hence
\be
M = \bQ_1 + \bQ_2 + \bQ_3\,.
\ee

\begin{figure}[ht!]
\begin{center}
\includegraphics[width=0.4\textwidth]{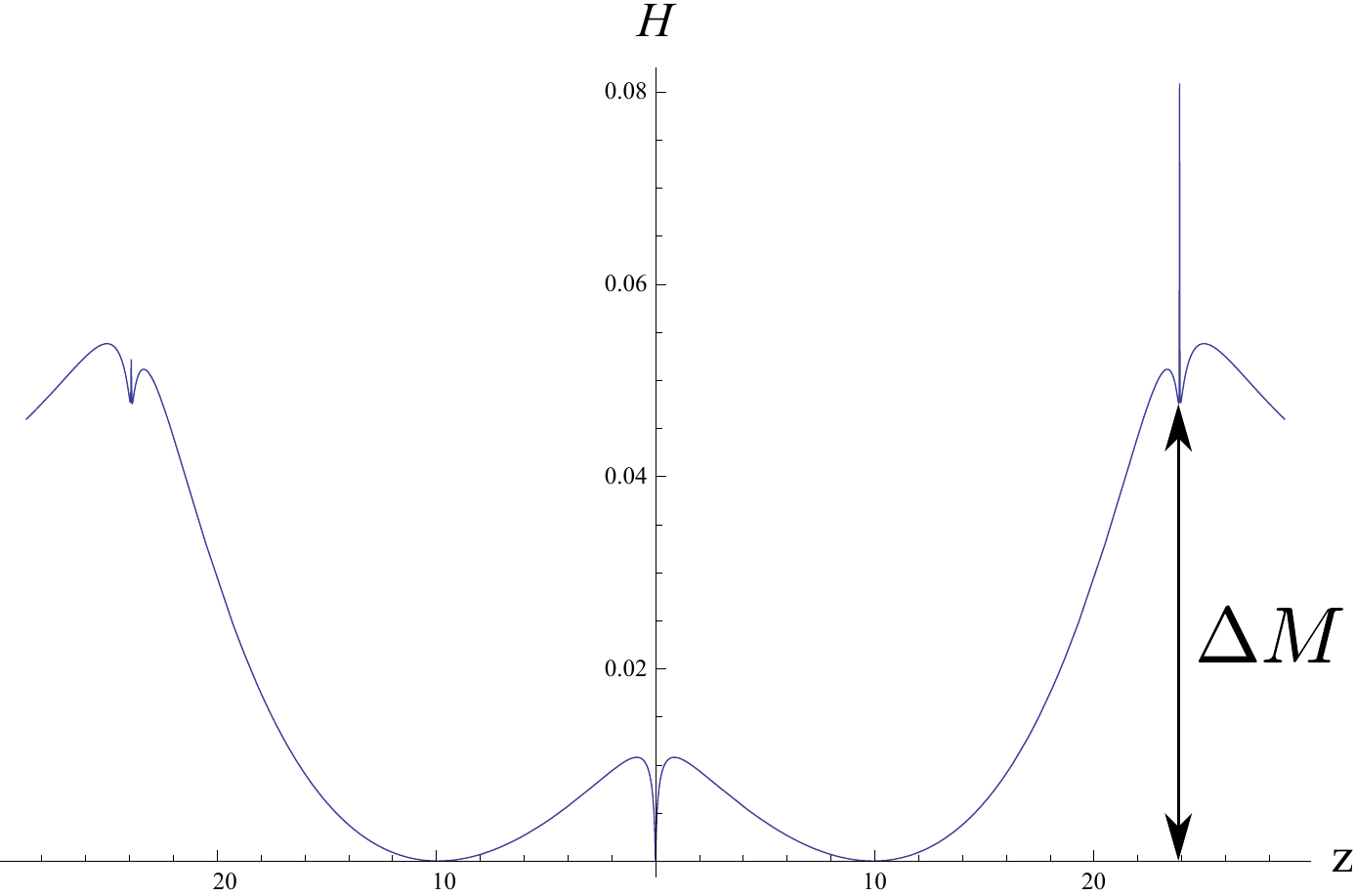}
\end{center}
\caption{The metastable supertube brings in an excess energy $\Delta M = M_{ADM}^{\rm BH} - \sum_I Q_I^{\rm BH}$.\label{fig:DeltaM}}
\end{figure}

Adding a supertube with charges $\pQ_1,\pQ_2$ increases the mass by
the value of the supertube potential at the minimum $\calh_{\rm min} =
\pQ_1 + \pQ_2 + \Delta M$ (see also Figure \ref{fig:DeltaM}). When the
minimum is supersymmetric $\Delta M =0$, and the resulting
configuration is a BPS microstate. When the supertube minimum is
metastable, the mass is: 
\be
M= \bQ_1 + \bQ_2 + \bQ_3 + \pQ_1 + \pQ_2 + \Delta M\,,
\ee
and the charges are
\be
Q_1^{\rm tot} =  \bQ_1 + \pQ_1\,, \qquad Q^{\rm tot}_2 = \bQ_2 + \pQ_2 \,\qquad Q^{\rm tot}_3 = \bQ_3\,.
\ee
Since, $M_{ADM} > \sum_I \bQ_I$, the configuration with a metastable
supertube has the charges and mass of a non-extremal black hole. The energy above extremality is exactly given by $\Delta M$:
\be
\Delta M = M - \sum_I Q_I^{\rm tot}\,.\label{eq:DeltaM_ExcessEnergy}
\ee

In appendix \ref{app:BlackHole}, we review the non-extremal rotating M2-M2-M2 black hole geometry. The solution depends on six parameters: a mass parameter $m$, three `boosts' $\delta_I$ and angular momentum parameters $a_1,a_2$ which are related to the ADM mass, charges and angular momenta
\eal{
M_{ADM}^{\rm BH} &=  \sum_I \frac m 4(e^{2\delta_I} + e^{-2\delta_I})\,, \qquad &J_1^{\rm BH} &=m( a_1 c_1 c_2 c_3 - a_2 s_1 s_2 s_3)\,,\\
Q_I^{\rm BH} &= \frac m 4 (e^{2\delta_I} - e^{-2\delta_I})\,, &J_2^{\rm BH} &=-m( a_2 c_1 c_2 c_3 - a_1 s_1 s_2 s_3)\,,
\label{eq:5dCharges}
}
where $c_I = \cosh \delta_I$ and $s_I = \sinh \delta_I$. We determine the parameters $\delta_I,a_i,m$ of the metastable state.  The parameters $\delta_I$ are given by the charges and parameter $m$ as
\be
\frac m 2 e^{2\delta_I} = Q^{\rm BH}_I + \sqrt{(Q^{\rm BH}_I)^2 + \frac{m^2}{4}}\,.\label{eq:Delta_I}
\ee
The parameter $m$ is determined by the energy of the metastable supertube as follows. With \eqref{eq:5dCharges} the energy above extremality \eqref{eq:DeltaM_ExcessEnergy} can be written in terms of $m$ and $\delta_I$ as
\be
\Delta M = \sum_I \frac m 2 e^{-2\delta_I}\,.
\ee
In the probe approximation, the supertube charges are small compared to those of the background. Then the non-extremal black hole is close to the supersymmetric limit ($m/Q^{\rm BH}_I \ll 1$) and the black hole charges are approximately those of the background and \eqref{eq:Delta_I} becomes
\be
\frac m 4 e^{2\delta_I} = \bQ_I\,.
\ee
The non-extremality parameter is then given by the charges and energy of the metastable state
\be
m = \sqrt{\frac{8 \Delta M}{\sum_I 1/\bQ_I}}\,.
\label{eq:mDeltaM}
\ee
In this approximation, the angular momentum parameters $a_1,a_2$  are:
\bea
J^{\rm BH}_L &\equiv& J^{\rm BH}_1 - J^{\rm BH}_2 = \frac {\sqrt{m}} 2 (a_1 +a_2)\sqrt{\bQ_1 \bQ_2 \bQ_3}\left( \frac 1 {\bQ_1} +\frac 1 {\bQ_2} +\frac 1 {\bQ_3} \right) \,,\nn
J^{\rm BH}_R &\equiv& J^{\rm BH}_1 + J^{\rm BH}_2 =  \frac{2}{\sqrt{m}}(a_1 -a_2)\sqrt{\bQ_1 \bQ_2 \bQ_3} \,.\label{eq:AngularMomentaBH}
\eea

\subsection{Comparing the microstates and the black hole}\label{ssub:Comparing}

As we explained in the Introduction, extremal black holes have an
infinite throat, and comparing the length of this throat to that of
the fuzzballs is meaningless. Comparing the thicknesses of the throats on the other hand gives automatically the same result: the thickness is only controlled
by the charges. For near-extremal black holes the thickness is also
largely controlled by the charges, so it will automatically be the
same for fuzzballs and black holes. On the other hand, non-extremal black
holes have a finite throat, and hence comparing the lengths of the
throats is now meaningful, and can indicate which fuzzballs are
expected to be more typical than the others, and whether fuzzballs differ
from the black hole away from the horizon microscopically or
macroscopically.

We denote the difference in the length of the non-extremal black hole
throat and that of its microstates by
\be
\Delta L \equiv L_{BH} - L_{MS}\,.
\ee
Although we have not backreacted the metastable bound state, we have
argued above that a small probe supertube will not significantly
change the geometry and hence $L_{MS}$ will be the length of the
supersymmetric microstate throat given by \eqref{11dgeometry}.  We can
estimate the throat length by integrating along the $z$-axis, from
the outermost center $z_{MS} \equiv z_7$ up to a suitable cutoff scale
$z_{\rm neck}$. The depth of the black hole throat is the metric
distance from the horizon at $\rho = \rho_+$ to the end of the throat
at $\rho = \rho_{\rm neck}$ with the metric \eqref{eq:BH_Metric}. A suitable cutoff is $z_{\rm
  neck}=\rho_{\rm neck} = (Q^{BH}_1 Q^{BH}_2 Q^{BH}_3)^{1/6}$.  The
expression for $\Delta L$ is quite complicated, but as we explain in
appendix \ref{app:Depth} we can make a very insightful approximation
through which we obtain
\be
\Delta L \approx \rho_{\rm neck} \ln \left(2\frac{\rho_{MS}}{\rho_{+}}\right)\,,
\ee
where we replaced the cutoff $z_{MS}$ by $\rho_{MS}$ in a spherically symmetric approximation of the microstate geometry.

Consider the following scaling of the supertube charges and of the
coordinates of the GH centers:
\bea
(\pQ_1,\pQ_2,d_3) &\to& e^\lambda (\pQ_1,\pQ_2, d_3)\,,\nonumber\\
\rho_{MS} &\to& e^{\mu} \rho_{MS}\,.\label{eq:scalings}
\eea         
The approximated difference in depths then goes as
\be
\boxed{\frac{\Delta L}{\rho_{\rm neck}} \to \frac{\Delta L}{\rho_{\rm neck}}- \frac{1}4\lambda+  \frac 3 4 \mu}\,.\label{eq:ThroatScaling}
\ee
This reveals that the black hole throat can be made deeper than that of the microstate ($\Delta L$ positive) by taking either smaller tubes or deeper background microstates.

To confirm this approximation, we evaluate $\Delta L = L_{BH} - L_{MS}$ numerically. We do this for supertubes of charges
\be
(\pQ_1,\pQ_2,d_3) = e^\lambda (10,-50,1)\,,
\ee
with $\lambda = -10,-9,\ldots,9,10$.  The supertubes are placed in the
nine  background   scaling  geometries   of  different  sizes   of  Table
\ref{tab:Merger}.  The  size of  the  black  hole  throat $L_{BH}$  is
calculated  from  \eqref{eq:BH_Depth}  for  the  rotating  black  hole
geometry \eqref{eq:BH_Metric}, and the  parameters of the black hole
are  extracted from  the  metastable supertube  minima  as in  Section
\ref{ssec:BlackHoleParemeters}. The  size of the  microstate throat is
obtained by integrating \eqref{eq:MS_depth}. We replace the background
microstate geometry by that of the extremal black hole.

\begin{figure}[ht!]
\begin{center}
 \subfigure[The difference in depths $\frac{\Delta L}{\rho_{\rm neck}}$ in terms of the tube charge scaling sizes $(\pQ_1,\pQ_2,d_3) = e^\lambda (10,-50,1) $ for the backgrounds 2, 6 and 9 of Table \ref{tab:Merger}.  \label{fig:Sizes}]{\boxed{\includegraphics[height=.24\textheight]{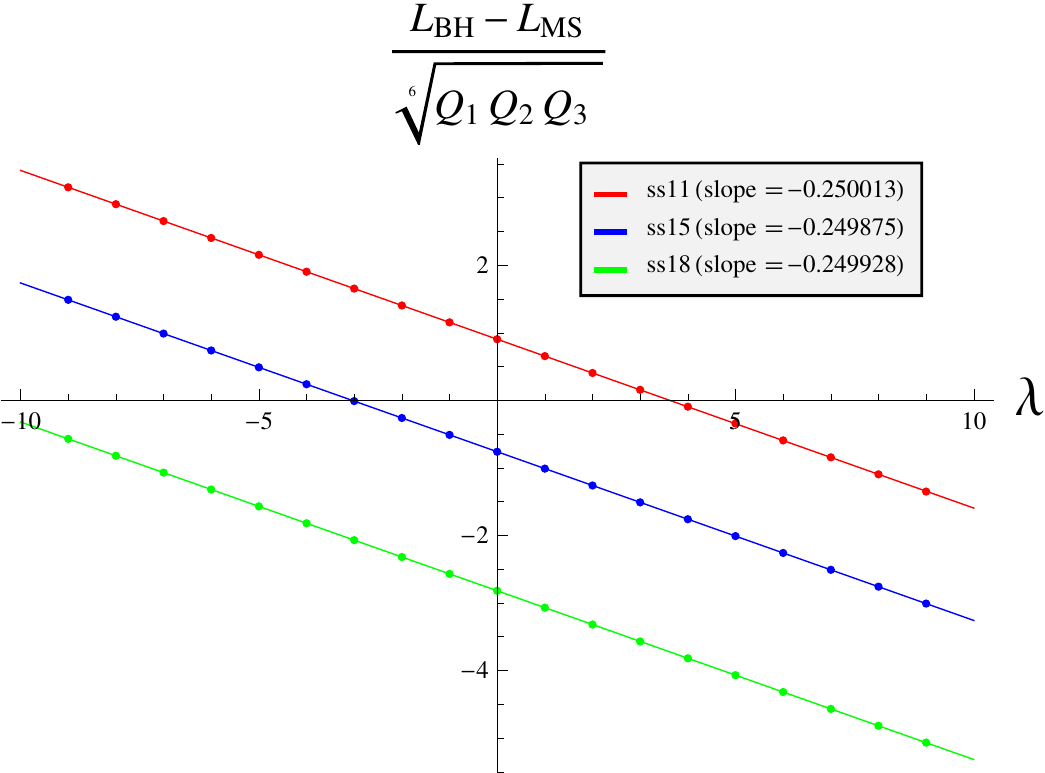}}}
\hspace{.01\textwidth}
\subfigure[The difference in depths $\frac{\Delta L}{\rho_{\rm neck}}$ in terms of the  scaling solution sizes measured by the logarithm of the position of the $7^{\rm th}$ center $z_7$.\label{fig:Backgrounds}]{\boxed{\includegraphics[height=.24\textheight]{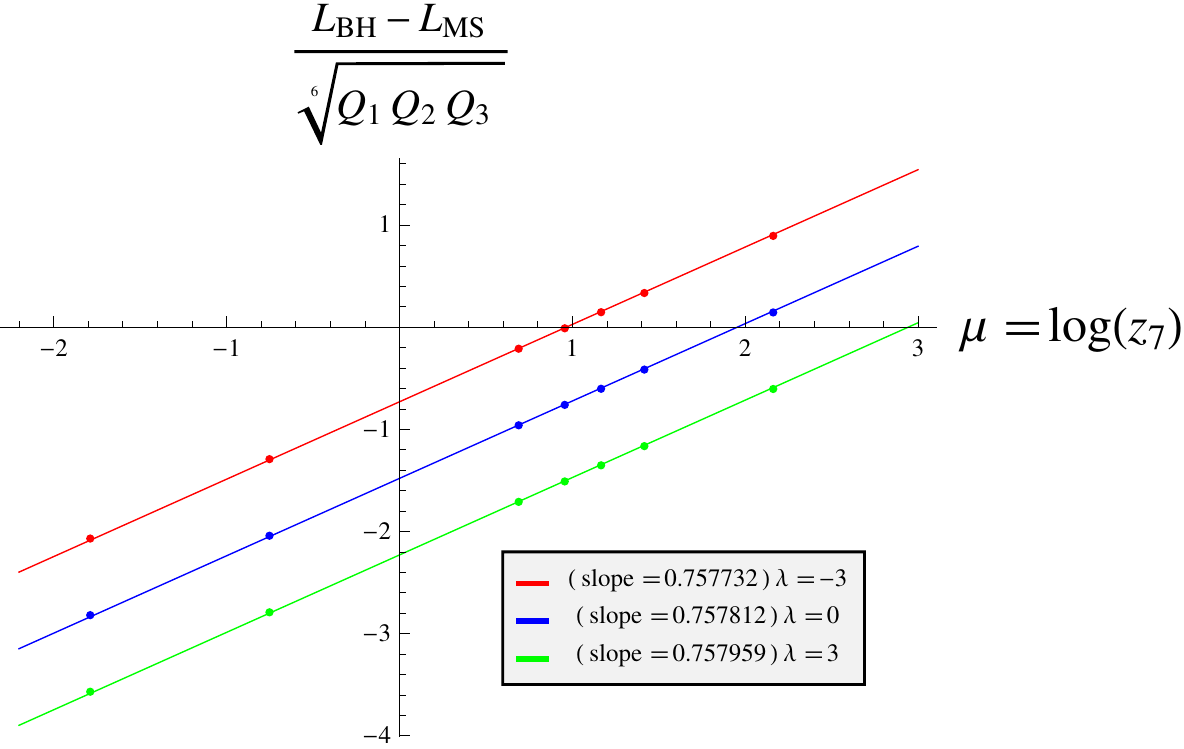}}}
\end{center}
\caption{The difference in depths $\Delta L = L_{BH} - L_{MS}$ for tubes of size $\lambda = -10,-9,\ldots,9,10$ in several scalings backgrounds.}
\end{figure}

We plot our findings in Figures \ref{fig:Sizes} and \ref{fig:Backgrounds}. In Figure \ref{fig:Sizes} we show the effect of scaling the tube charges. We plot $\Delta L$ for tubes of various sizes ($\lambda = -10,-9,\ldots,9,10$), in three scaling solutions of Table \ref{tab:Merger}.  We find all possibilities: microstates that are deeper, of the same depth, and shallower than the black hole. By making the tubes smaller, the black hole can always be made deeper than the microstate. It is also clear that $\Delta L$ has the scaling behavior anticipated in \eqref{eq:ThroatScaling}.

In Figure \ref{fig:Backgrounds} we show the effect of putting the tube
in backgrounds of different scaling size and depth. We plot $\Delta L$
for tubes in all nine scaling solutions (2,6 and 9 of Table
\ref{tab:Merger}), in terms of $\mu \equiv \log z_7$, where $z_7$ is
the position of the outermost center of the scaling background. This
reveals that the approximately linear scaling \eqref{eq:ThroatScaling}
still holds.

\subsection{The range of validity of our construction}

Having obtained non-extremal microstates by placing probe supertubes
inside long supersymmetric fuzzballs, it is important to study the
ranges of charges in which our construction is valid. Clearly, since
we have not backreacted the supertube, and treated them as probes, we
are automatically assuming that their charges and dipole charges are
much smaller than those of the GH centers and we  are only describing configurations whose mass above extremality is
much smaller than the sum of the charges. These correspond to
microstates of near-extremal black holes.

A possible mechanism for invalidating our construction is if the
microstates we create will have closed timelike curves. In the absence
of backreaction one cannot say precisely when or whether this will
happen; however, one can estimate whether the angular momentum of a
microstate is larger or smaller than the angular momentum that would
cause closed timelike curves in a black hole of identical charges and
length. 

The non-extremal black hole geometry has closed timelike curves  unless $m \geq(a_1 \pm a_2)^2$. This gives two
`cosmic censorship bounds'  on the black hole angular momenta. For a
near-extremal black hole, the angular momenta
\eqref{eq:AngularMomentaBH} have to satisfy:
\bea
J^{\rm BH}_L &\leq& \frac {m} 2 \sqrt{\bQ_1 \bQ_2 \bQ_3}\left( \frac 1 {\bQ_1} +\frac 1 {\bQ_2} +\frac 1 {\bQ_3} \right) \,,\nn
J^{\rm BH}_R &\leq&  2 \sqrt{\bQ_1 \bQ_2 \bQ_3} \,.\label{eq:AngularMomentaCC}
\eea
The second bound is automatically satisfied because we are starting
with a BPS microstate of a black hole with a large horizon area, and
adding a probe supertube only changes the charges and $J_R$ by very
small amounts.

The first bound is more problematic. Since the BPS microstate has
$J_L=0$, the resulting metastable microstate will get its left-moving
angular momentum entirely from the $\vec E \times \vec B$ interactions
between the supertube and the background. If we call this contribution
$J_L^{\rm tube}$ and express the parameter $m$ in terms of the energy
of the metastable supertube $\Delta M$, this bound becomes
\be
J_L^{\rm tube} \leq A(Q)\sqrt{ \Delta M}\,, \qquad A(Q) = \sqrt{2}\sqrt{\bQ_1 \bQ_2 \bQ_3}\sqrt{ \frac 1 {\bQ_1} +\frac 1 {\bQ_2} +\frac 1 {\bQ_3} }\,.
\ee
Since both the angular momentum $J_L^{\rm tube}$ and $\Delta M$ scale
linearly with the tube charges, we see that in a solution of fixed
length this condition will be violated when the tube charges become very
large.

Alternatively, one can
consider a supertube with fixed charges in a solution whose length is
dialed by hand by bringing the centers together on the GH base. The
mass above extremality $\Delta M$ is linear in the inter-center
separation, while we expect (from the known supersymmetric solutions)
that the $\vec E \times \vec B$ interactions that give rise to
$J_L^{\rm tube}$ will remain constant. Hence, a solution with a
single supertube that becomes too deep will start having charges and
angular momenta outside of the cosmic censorship bound, and will most
likely have closed timelike curves.

Of course, the way to avoid all these complications is to use the fact
that the original solution is $\mathbb{Z}_2$-symmetric and place two identical
supertubes in metastable minima symmetric around the origin, such that
resulting configuration preserves this symmetry.  In such a  symmetric
configuration the contribution to $J_L^{\rm tube}$ from the
interaction of the supertubes with the background vanishes, and the
cosmic censorship conditions are always satisfied.

\subsection{The force on probe branes}

Another quantity that one can compute in both the near-extremal black
hole geometry and in the non-extremal fuzzballs we construct is the
force on a probe brane whose charge is carried by the black hole.
Of course, this force is identically zero in the BPS black hole and in
the BPS fuzzballs, but now we have more mass than charge, and we
expect such a probe brane to start feeling a force.

The potential for an M2 brane wrapping the $I^{\rm th}$ torus
$T_I^2$, in the Cvetic-Youm black hole (appendix \ref{app:BlackHole})
is easy to compute\footnote{The force is $\vec F_{BH}=q_{M2}\frac{\partial V_{BH}}{\partial \vec r_{M2}}$.}
\be
V_{BH}^{(I)}= V_{DBI} + V_{WZ} = \frac{1}{H_I}(\sqrt{H_m}-\coth{\delta_I})\,.
\ee
As before, in the near-extremal limit with no rotation
$Q_I^{BH}\approx Q_I$ and  $m \ll Q_1$.
The leading-order contribution to the potential of a probe M2 brane
wrapping the torus $T_I^2$ is proportional to what we may define as the mass above extremality in the $I^{\rm th}$ channel:
\be
V_{BH}^{(I)}=\frac{\Delta M_I}{\rho^2}\, , \qquad \text{where} \quad
\Delta M_I \equiv \frac{m^2}{8Q_I}  \qquad \text{and} \quad
\Delta M = \sum_{I=1}^3 \Delta M_I +\mathcal{O}(m^3)\,.
\label{eq:Vbh}
\ee
Note that probe branes wrapping different tori will correspond upon
compactification to five dimensions to point particles with different
types 
of U(1) charges, and feel different forces.

To compute the force on a probe M2 brane in the non-extremal fuzzball
one may think naively that one needs to construct the fully
backreacted solution corresponding to metastable supertubes, but this is not so.
Using the fact that in the absence of a metastable supertube one can add a BPS M2 brane at a large
distance away from the throat without breaking 
supersymmetry, one can calculate the action of a metastable supertube in a microstate geometry both with and without the brane. The difference between the two
actions gives then by Newton's third law the potential felt by far-away M2 brane
as a function of its position, which can then be used to determine the
force it feels. In the examples where a backreacted solution exists,
this method,
first introduced in KKLMMT \cite{Kachru:2003sx}, reproduces correctly the
force computed from supergravity
\cite{Bena:2010ze,Dymarsky:2011pm,Bena:2011hz,Bena:2011wh}.

Adding an M2 brane with charge $q_{M2}$ far away from the scaling centers introduces
another term in the M2 harmonic function 
\be
L_I=\sum_{i=1}^7 \frac{\ell_i}{|\vec{r}-\vec{r}_i|}+\frac{q_{M2}}{|\vec{r}-\vec{r}_{M2}|}\,,
\ee
and for small charges this changes the energy of the metastable
supertube by  
\be
V_{MS}^{(I)} = \frac{\partial \mathcal{H}}{\partial L_I}\bigg|_{\rm min} \frac 1 {|\vec{r}_{\rm min}-\vec{r}_{M2}|}\,,\label{eq:potentialMS}
\ee
which by Newton's third law gives then the potential felt by the M2 brane
in  the non-extremal fuzzball. 

Given that our non-extremal microstates have the same mass and charges
as a non-extremal black hole, we would expect by Birkhoff's theorem
that the leading-order term in the potential felt by an M2 brane far
away from the region of the throat would be the same. However, the
leading-order term in \eqref{eq:potentialMS} is not of the same form
as \eqref{eq:Vbh}; in particular the microstate attractive
potential \eqref{eq:potentialMS} does not scale properly with the length of the microstate throat:
\be
\frac{V_{MS}^{(I)}}{V_{BH}^{(I)}} \sim \frac1 {L_I}\,.
\label{eq:scaleratio}
\ee
Since $1/L_I$ is linear in the inter-center distances of the scaling
background, the force with which the microstate attracts the M2 brane
vanishes as one considers deeper and deeper microstates with the same
mass. Furthermore, another surprise is in store. A microstate with a
metastable supertube with electric charges $q_1$ and $q_2$ will
attract M2 branes with charges $Q_1$ and $Q_2$, but will repel M2
branes with charge $Q_3$. This can be seen both by investigating
\eqref{eq:potentialMS}, or by evaluating the potential numerically:
\be
\frac{V_{MS}^{(1)}}{V_{BH}^{(1)}} \approx 3.0\times 10^{-5} \,z_6\,,\qquad \frac{V_{MS}^{(2)}}{V_{BH}^{(2)}} \approx 2.9\times 10^{-5}\,z_6\,,\qquad \frac{V_{MS}^{(3)}}{V_{BH}^{(3)}} \approx -6.4 \times 10^{-5}\,z_6\,.
\ee
The ``wrong'' sign of $V_{MS}^{(3)} $ and the linear dependence of these
ratios on $z_6$ implies that one cannot hope to obtain the ``correct''
black-hole force on probe M2 branes from microstates constructed
this way. One can also imagine constructing other types of black hole
microstates, by placing single anti M2-branes or other more complicated objects inside bubbling
geometries; however, the ``force problem'' persists, and it can be summarized as
follows:

Given a background that does not attract M2 branes, the mass above
extremality generated by adding a probe that breaks supersymmetry
inside a long throat generically goes like the mass of that probe
divided by the warp factor at the bottom of the throat. On the other
hand, if one computes ``\`a la KKLMMT'' the force on a probe M2 brane,
this force scales generically like the mass of the probe divided by
the \emph{square} of the warp factor, and hence like the ADM mass of the
solution divided by the warp factor at the bottom. If one now makes
the throat longer or shorter keeping the mass fixed, the force in a
microstate changes, unlike in a black hole solution where this force
is always proportional to the mass above extremality.

This force analysis indicates that the non-extremal microstates
obtained by placing single metastable supertubes inside BPS
microstates do not attract M2 branes in the way one may naively expect of a typical black hole 
microstate. The underlying reason for this is that the supertube couples
not only to the warp factor and electric fields but also to extra
scalars in five dimensions, which come from the volume moduli of the
torus. This extra interaction, which is absent for M2 probes in the
background of the black hole, leads to the different scaling behavior
of the force on a probe M2 in the microstate background and to the
repulsive force felt by  an M2-brane along the third torus.

Even if the microstates we obtain by
placing one metastable supertube do not attract M2 branes the way the black hole 
does,  one can clearly bypass this problem and construct very large numbers
of microstates that attract M2 branes typically by placing several
species of metastable supertubes, and by fixing the supertube charges
such that the length of the microstate is exactly that of the black
hole \eqref{eq:ThroatScaling} and furthermore such that at this length the forces are
exactly those of the black hole \eqref{eq:potentialMS}.
However, this is more a matter of engineering, and can
obscure an important piece of physics that the force
computation reveals: the fact that the force on a probe M2 branes
varies wildly from microstate to microstate, and can be even negative,
implies that M2 branes will feel the thermal fluctuations between
various fuzzballs as thermal fluctuations in the force even if they
are quite far away from the black hole. Thus, these M2 branes will
become aware of the existence of fuzzballs and of the breakdown of
classical physics further away from the horizon than other probes.

\section{Fuzzballs of Fire or Fuzzballs of Fuzz  (in lieu of Conclusions)}
\label{ch:conclusions}

In this paper we have used probe supertubes to construct microstate geometries, or fuzzballs, that have the same mass and charges as three-charge non-extremal black holes. We computed the length of the throats of these solutions, and found that one can easily build microstates whose throats are longer,  shorter or have the same length as the throat of the black hole. Since in our construction there is no dynamical mechanism that sets the microstate length to be the same as that of the black hole throat, this indicates that this mechanism may be entropic: there will be many more microstates of black hole throat length than shorter or longer ones. Of course, to produce such an entropic argument one needs first to make sure that our method for constructing non-extremal microstates can produce at least a subset of the typical ones, and them count these microstates.

However, the absence of a dynamical mechanism for fixing the microstate length indicates that fuzzballs will differ from the black hole at macroscopic distances from the horizon, and not just in its vicinity (as recently mentioned as a possibility in \cite{Almheiri:2012rt}).
Of course, this intuition applies to near-extremal fuzzballs, and the extrapolation to more generic black holes may break down. Nevertheless, one can use near-extremal black holes as a testing ground for all the ideas proposed in relation to infalling observer physics, firewalls, black hole complementarity and spacelike singularity resolution and this is the purpose of this section.

\subsection{What does an in-falling observer see ?}
The first question one can try to address is the scale at which an infalling observer stops experiencing spacetime. In the most straightforward interpretation of the fuzzball proposal, which one may call a ``fuzzball of fire'' interpretation, the classical geometry breaks down at the horizon, and is replaced by an ensemble of fuzzballs \cite{Bena:2007kg}. Hence, the horizon is the scale where the ``thermodynamic'' description of physics (in terms of a classical spacetime) breaks down, and the ``statistical'' description (in terms of fuzzballs) takes over. In a naive analogy with an ideal gas, the scale of the horizon is like that of the mean free path, and hence we might expect the incoming observers to experience large statistical fluctuations in the same way in which a particle of smaller and smaller size in a gas experiences larger and larger fluctuations. Below a certain size of the particle the Brownian motion deviations overtake the classical trajectory, and the notion of ``particle moving in a continuous fluid'' breaks down. In the same way, a particle far away from a black hole experiences a classical spacetime, but as the particle approaches the horizon the statistical fluctuations become stronger and stronger, and at the horizon the notion of ``particle moving in a classical spacetime'' breaks down. 

From the point of view of the incoming particle, the increasing fuzzball fluctuations it feels at the scale of the horizon are not a very pleasant experience, which agrees with the recent proposal of \cite{Almheiri:2012rt} that an incoming particle must see a firewall at the horizon scale in order for the information paradox to be solved. In \cite{Chowdhury:2012tr} one of the authors and Chowdhury have argued that the pleasantness an in-falling particle experiences depends on its energy; particles of the order of the Hawking radiation should thermalize in the bath of out-going radiation which thus constitutes a firewall for these in-falling particles, while particles much heavier than Hawking radiation should pass the bath nearly unaltered. 

Recently Mathur has argued for a new approximate complementary for observers heavier than Hawking radiation falling into a fuzzball. According to this ``fuzzball complementarity'' paradigm \cite{Mathur:2012zp,Mathur:2012dx,Mathur:2012jk}, which draws from recent ideas of \cite{VanRaamsdonk:2009ar,VanRaamsdonk:2010pw}, the scale of {fuzzball thermalization/loss of spacetime} experience does not depend only on the location of the infalling observer, but also on its energy. Heavier observers will continue experiencing a spacetime even after they have passed the horizon scale and have entered the fuzzball region, and for these observers spacetime will emerge from the quantum superposition of the fuzzballs. 
On the other hand lighter particles, of order the Hawking radiation energy, stop experiencing a classical spacetime at the horizon -- this is necessary in order for the Hawking radiation to be able to carry off the black hole information to infinity and solve the information paradox.
According to the ``fuzzball complementarity'' paradigm, the analogy with the ideal gas in the ``fuzzball of fire'' argument above is not so straightforward, essentially because in the ideal gas there is only one scale, while when describing an observer falling into a black hole one has two scales: the observer mass and its location.

It is hard to tell directly whether our non-extremal fuzzballs will be felt by an incoming observer as fuzzballs of fire or as fuzzballs of fuzz (in the sense of fuzzball complementarity). To do this one would have to construct first more generic non-extremal fuzzballs, and then to scatter various particles off them. Such a research programme is feasible; one can in particular use the quiver quantum mechanics that describes these fuzzballs in the regime of parameters where gravity is turned off \cite{Denef:2002ru}, and analyze the scattering of various charged centers, as one does in supergoop studies \cite{Anninos:2012gk}. One can analyze for example the collision of a multicenter near-extremal fuzzball goop with a center whose charges are much bigger than those of the centers that compose the fuzzball, and see whether such a center gets absorbed by the fuzzball goop as soon as it reaches it or traverses it with impunity, and if so how does the trajectory of this center differ from the trajectory in a single-center black hole geometry.

However, even before such a calculation is done, there are two features of our construction that are relevant in the fuzzballs of fire/fuzz discussion. First, the fact that nothing dramatic happens as the throat of microstate geometries becomes longer or shorter than that of a black hole implies that the difference between fuzzballs and the classical black hole is not strongly suppressed immediately above the horizon scale. Hence, if an observer heavier than Hawking particles continues to experience a spacetime below the horizon scale then, by extension, an observer lighter than the Hawking particles should stop experiencing a spacetime {\em above} the horizon scale. The physics of this possibility can get quite unpalatable: a spaceship orbiting at say five Schwarzschild radii above the horizon cannot send to another nearby spaceship any photons that have an energy lower than the Hawking radiation energy divided by five to some power; such photons are thermalized by the ensemble of fuzzballs already at that scale and hence cannot propagate. 

Since we do not expect observers, be they very small, to start experiencing large statistical fluctuations of spacetime and dissolve in the fuzz far away from the horizon, this Gedanken experiment seems to tilt the balance against ``fuzzball complementarity'', and towards the ``fuzzball of fire'' interpretation. On the other hand, another piece of fuzzball physics in our construction seems to incline the balance backwards:

We have computed the force on probe M2 branes (that experience no force when the fuzzball is supersymmetric), and we have found that this force varies wildly, and can even change sign when one goes from one metastable fuzzball to another. It may be that this wild variations in the force are just a feature of the very specific type of fuzzballs we have succeeded to construct, and some yet-to-be-constructed more typical microstates  will not attract M2 branes in such erratic ways. However, it is also possible that we have uncovered a fundamental feature of fuzzballs of near-extremal black holes: they may attract very erratically the components of the extremal black hole with the same charges. 

Now, if these very special probes see the thermal noise from the fuzzball and experience statistical fluctuation already  at a very large distance, way above the horizon scale, it does not seem so far-fetched that other observers, in particular those with energy below that of Hawking radiation, could also see this thermal noise far away from the horizon, while other (more massive) observers continue experiencing a spacetime well into the fuzzball, as the ``fuzzball complementarity'' paradigm indicates.

\subsection{Is Hawking radiation coming from brane-flux annihilation ?}

Besides being the first examples of non-extremal fuzzballs and maybe a way to realize firewalls in string theory, one can also use our configurations to explore other pieces of black hole physics. The first is Hawking radiation. The metastable supertubes we use to construct the fuzzballs decay into supersymmetric vacua via brane-flux annihilation \cite{Kachru:2002gs,Bena:2011fc} and this decay corresponds to the near-extremal black hole emitting its last Hawking radiation quantum and becoming an extremal black hole. This process is quite difficult to study from the black hole side, essentially because thermodynamics breaks down \cite{Preskill:1991tb,Kraus:1994fj}. A comparison of the fuzzball decay rates (which one can compute rather straightforwardly) to the near-extremal black hole emission rate may shed light on how thermodynamics breaks down, and also on which fuzzballs are more typical than others: the decay rates depend on how big the bubbles of a fuzzball are, and can be used to determine the typical bubble size.

The other important question is what is the backreacted solution corresponding to our metastable supertubes. The JMart solution \cite{Jejjala:2005yu}, which is one of the two known fully-backreacted non-extremal fuzzballs, is known to have ergoregions, and its instability  \cite{Cardoso:2005gj} has been argued \cite{Avery:2009tu} to correspond to Hawking radiation. The other fully-backreacted non-extremal fuzzball, the running-Bolt solution \cite{Bena:2009qv}, is also unstable, 
but its instability does not come from ergoregions. Our configurations on the other hand are metastable, at least in the probe approximation. It might be possible that the energy of some metastable supertubes will decrease by taking one of the GH centers off the symmetry axis, and if this happens the fully backreacted solution will probably be unstable as well. However, it is also very likely that one will be able to construct metastable supertubes that remain metastable when fully backreacted, and thus will have very different physics from the JMaRT and running-Bolt solutions.

The other issue with the metastable supertube backreaction is that most of the work analyzing the backreaction of antibranes in backgrounds with charge dissolved in fluxes indicates that such backreacted solutions have unphysically-looking singularities, both at first order \cite{Bena:2009xk, Bena:2010gs, Bena:2011hz, Giecold:2011gw, Dymarsky:2011pm,Massai:2012jn} and when looking at the fully-backreacted solutions \cite{Blaback:2011nz,Bena:2012bk}. If one would naively extend this result to our work, one might expect that anti-M2 branes in long BPS throats will also
give rise to unphysically-looking singularities. However, in our construction we are not using bare antibranes, but supertubes that carry two kinds of antibrane charges, and also have a dipole charge and a nonzero angular momentum. The advantage of supertubes is that if they are solutions of the DBI Hamiltonian they backreact into geometries that are smooth in the D1-D5-P duality frame  \cite{ Bena:2008dw}. Hence we expect the backreaction of our metastable supertubes to give rise to regular solutions; it would be very interesting to confirm this by constructing directly this challenging non-supersymmetric cohomogeneity-two solution. 

\subsection{Are spacelike singularities resolved backwards in time ? }

The existence of microstate solutions that have the same mass, charges and throat length as  non-extremal black holes indicates that the singularity of these black holes will most likely not only be resolved to the inner horizon (as one may expect by extrapolating the extremal black hole result) but all the way to the outer one, which is backwards in time from where the singularity is, as the Penrose diagrams in Figure \ref{fig:singresol} show. One can now try to see what this intuition may tell us about singularity resolution.

Indeed, since the Penrose diagram of the near-extremal black holes is the same as that of all Reissner-Nordstr\"{o}m black holes, one can extrapolate our result and assume that the singularity of all Reissner-Nordstr\"{o}m black holes is resolved all the way to the outer horizon. One can then take the small charge limit (in which the inner horizon and the timelike singularity merge to form a spacelike singularity) and infer that the spacelike singularity of the zero-charge Schwarzschild black hole is also resolved backwards in time, all the way to the horizon.

If this is indeed the correct pattern of the resolution of spacelike singularities in string theory, one can ask two questions: 
\begin{enumerate}
 \item What is the mechanism by which this happens and the corresponding scale?
 \item What does this imply for the physics of other spacelike singularities, like the cosmological ones?
\end{enumerate}

Both questions have several possible answers, and we leave it to the reader who is unhappy with them to find more compelling ones. To answer Question 1, one can always argue that singularities in string theory have low-mass degrees of freedom, that destroy the spacetime on macroscopic distances. One can also refine the answer, and argue \cite{Mathur:2008kg} that an incoming shell that will form such a singularity in the future will enter in a region where there are a very large number of fuzzball-like states, and even if the probability of tunneling into any of them is tiny, since there are so many of them, the incoming shell will tunnel in the fuzz with probability one. The size of the region where the singularity is resolved depends on the mass of the singularity, and on the density of fuzzballs.

The answer to the second question depends largely of the mechanism for the backwards in time resolution. If this mechanism involves tunneling into fuzzball-like configurations that live near say a Big Crunch singularity, then the scale for the resolution will probably be given by the typical size of these configurations (which is quite hard to estimate at this point, in the absence of any explicit Big-Crunch-resolving fuzzballs). Furthermore, one can argue that such configurations were also present in the early universe, where they can again be thought of as giving a ``forward in time'' resolution of the Big Bang singularity, and their physics might have cosmological implications \cite{Chowdhury:2006pk,Mathur:2012vb}. We leave the fascinating exploration of these possibilities to future work.

\section*{Acknowledgements}

We would like to thank Borun D. Chowdhury, Samir Mathur, Thomas Van Riet and Nick Warner for useful discussions. 
This work was supported in part by the ANR grant 08-JCJC-0001-0 and by the ERC Starting Independent Researcher Grant 240210 - String-QCD-BH. BV would like to thank Evelien Dejonghe for her  support. 

\appendix{

\section{Smooth scaling backgrounds}
\label{app:3chargescaling}

We focus on three-charge backgrounds that are microstate geometries of black holes and black rings. Microstate geometries are everywhere smooth and free of horizons, such that each individual geometry carries no entropy. They have the same mass, charges and angular momenta as their black hole or black ring counterpart. Deep microstate geometries have a scaling behavior: the centers can be put arbitrarily close such that the geometry develops a very long throat, while the curvature is small everywhere.

\subsection{Smoothness and regularity.} 

The first physical requirement on the background is the absence of closed timelike curves (CTC's) in the geometry, giving the necessary conditions:\footnote{The sufficient no-CTC condition, which insures the existence of a time function is $Z_1Z_2Z_3 V - \mu^2V^2 \geq\ \omega^2$ \cite{Berglund:2005vb}.}
\eqn{
Z_1Z_2Z_3 V - \mu^2V^2\geq0\quad \text{and} \qquad
V Z_I \geq0\,.\label{eq:CTC_cond}
}
Note that this ensures that the potential \eqref{eq:Hamiltonian} is well-defined (radicand under square root is positive).
To have a smooth geometry, the warp factors and the function $\mu$ appearing in the angular momentum one-form $k$ must be regular at the sources of the harmonic functions. This yields relations between the charges and the constants in the harmonic functions. The constants are further constrained by demanding asymptotic flatness. We will choose the harmonic functions $V$ (Taub-NUT charges) and $K^I$ (dipole charges) to be fixed as:
\eqn{ 
V=\sum_{j=1}^N \frac{v_j}{r_j}\,,\qquad K^I=\sum_{j=1}^N \frac{k_j^I}{r_j}\,.
}
Regularity requires these harmonic functions to be sourced at the same points and one must take $v_j \in \mathbb{Z}$. For the base metric to be asymptotically $\mathbb{R}^4$ one must impose
\eqn{ 
\sum_{j=1}^N v_j=1.
}
Then smoothness determines $L_I$ (M2 charges) and $M$ (momentum along $\psi$) to be
\eqn{
L_I=1-\frac{1}{2}C_{IJK} \sum_{j=1}^{N} \frac{k_j^J k_j^K}{v_j} \frac{1}{r_j}, \qquad M=m_0+\frac{1}{12} C_{IJK} \sum_{j=1}^N \frac{k_j^I k_j^J k_j^K}{v_j^2} \frac{1}{r_j},
}
with
\eqn{
m_0=-\frac{1}{2} \frac{\sum_{j=1}^N \sum_{I=1}^3 k_j^I}{\sum_{i=1}^N v_i} = -\frac{1}{2} \sum_{j=1}^N \sum_{I=1}^3 k_j^I.
}
After imposing regularity and smoothness as well as asymptotic flatness there is a residual freedom in choosing $N-1$ Taub-NUT charges and $N$ dipole charges.

The microstates are `bubbled' geometries. For $N$ centers, there are $N-1$ non-trivial two-cycles, or bubbles, on the GH base. The cycles are supported by $N-1$ non-trivial fluxes $\Pi^{(I)}_{ij}$:  
\eqn{ 
\Pi_{ij}^{(I)}\equiv \frac{K^I}{V}\Big|_{r_j}-\frac{K^I}{V}\Big|_{r_i} = \Big( \frac{k^I_j}{v_j}-\frac{k^I_i}{v_i}\Big).\label{eq:smoothbubble}
}
We depict such a geometry in Figure \ref{fig:Bubbling}.

\begin{figure}[ht!]
\begin{center}
\includegraphics[width=.5\textwidth]{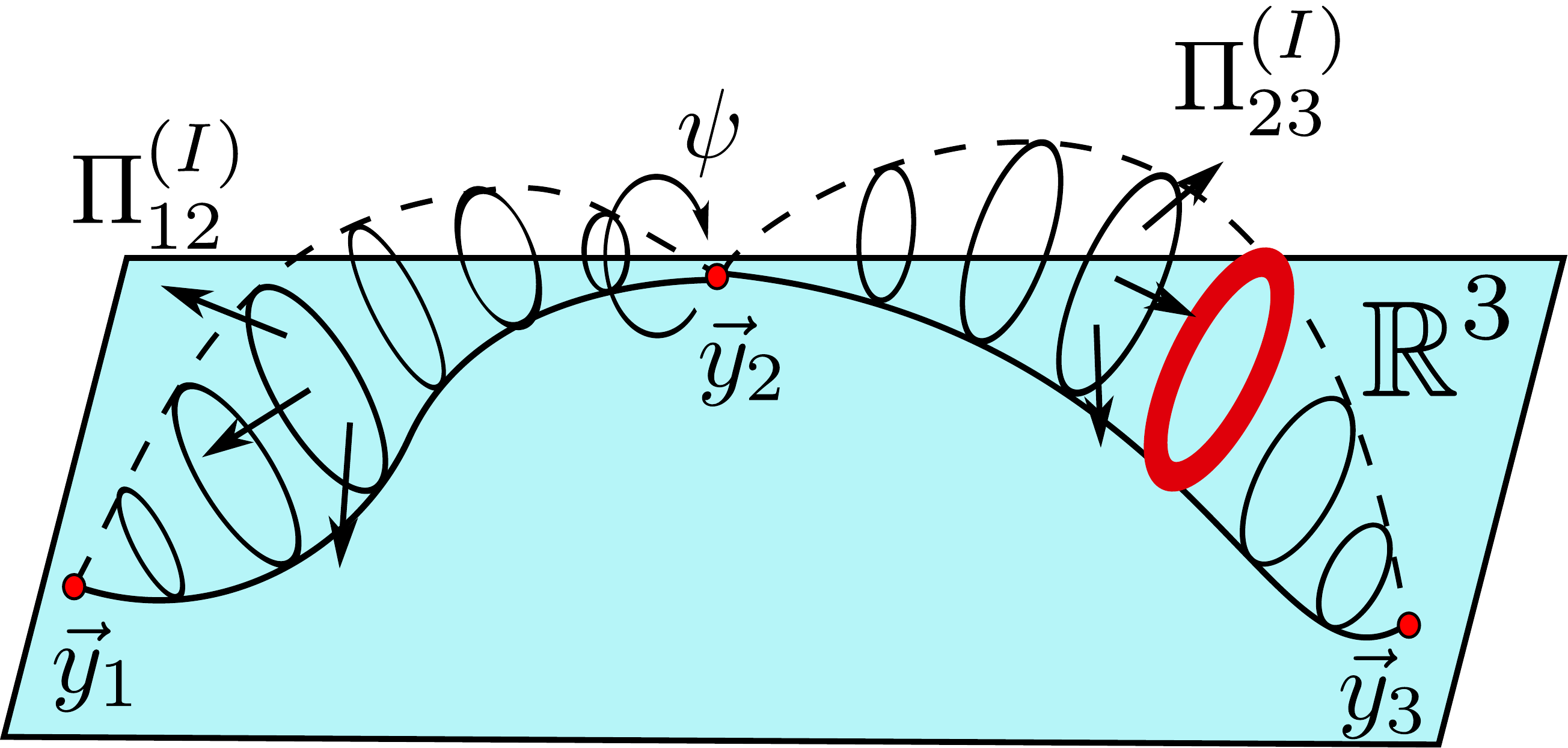}  
\end{center}
\caption{Smooth three-charge bubbling geometry with a supertube (red) placed on one of the cycles
along $\psi$.\label{fig:Bubbling}}
\end{figure}

The smoothness condition together with the first condition of (\ref{eq:CTC_cond}) leads to a further requirement that ensures the absence of CTC's: $\mu$ has to vanish at each center, since for $r_i\to 0$ the $Z_I$'s tend to finite values while $V^{-1}$ goes to zero. This gives $N-1$ \emph{bubble equations} \cite{Bena:2005va,Berglund:2005vb,Saxena:2005uk,Bena:2007kg}. By writing the charges and constants in the harmonic functions as vectors $\Gamma_i = (v_i,k_i^I,\ell_{I,i},m_i)$ and $h = (V_\infty,K^I_\infty,L_{I,\infty},M_\infty)$ these are:\footnote{In more general solutions \cite{Denef:2000nb,Denef:2002ru,Bates:2003vx} these equations come from imposing that $\omega$ should have no Dirac-Misner strings at the centers, but in smooth backgrounds this is equivalent to \eqref{eq:bubble_quantized}.}
\be
\forall i:\qquad \sum\limits_{\substack{j=1\\j\neq i}}^N \frac{\langle \Gamma_i, \Gamma_j\rangle}{r_{ij}}  + \langle \Gamma_i,h\rangle= 0\,,\label{eq:bubble_quantized}
\ee
where $\langle \Gamma_i, \Gamma_j\rangle = v_i m_j - m_i v_j + \tfrac 12 (k_i^I \ell_{I,j} - \ell_{I,i}k^I_j) $ is the symplectic product and $r_{ij}=|\vec{r}_j-\vec{r}_i|$ are the inter-center distances.
 For smooth solutions, the bubble equations can be written in terms of the magnetic two-form fluxes $\Pi_{ij}^{(I)}$ through the bubbles as:
\eqn{ 
\frac{1}{6} C_{IJK} \sum_{\stackrel{j=1}{j\neq i}}^N \Pi_{ij}^{(I)} \Pi_{ij}^{(J)} \Pi_{ij}^{(K)} \frac{v_i v_j}{r_{ij}} = -2 \Big(m_0 v_i + \frac{1}{2} \sum_{I=1}^3 k_i^I\Big).
\label{eq:bubbleeq}
}
The bubble equations relate the magnetic flux through each bubble to the physical size of each bubble.

\subsection{Asymptotic charges and angular momenta}

We give the charges and angular momenta of the five-dimensional solutions. Since the solution is invariant under the gauge transformation $K^I \rightarrow K^I + c^I V$, or $k^I_j + c^I v_j$, for any constant $c^I$, we define asymptotic quantities in terms of the gauge invariant flux parameters:
\eqn{ 
\kt^I_j \equiv k^I_j - v_j\left(\sum_{j=1}^N k^I_j\right)\,.
}
The electric charges of the solution as measured at infinity are extracted from the $\rho^{-2}$ term, with $r=\frac{1}{4} \rho^2$ in the expansion of the warp factors $Z_I$:\footnote{To isolate the charges of the solution one needs to take (\ref{eq:GH}) to a standard polar form for $\mathbb{R}^4$ via $r=\frac{1}{4} \rho^2$.}
\eqn{ 
\bQ_I = -2 C_{IJK} \sum_{j=1}^N \frac{\kt^J_j \kt^K_j}{v_j} .
\label{eq:elcharges}
}
In five dimensions there are two angular momenta, which are read off from the asymptotic behavior of $k$ in (\ref{eq:warprot}) from the terms that have a $\rho^{-2}$ fall-off:
\eqn{ 
k \sim \frac{1}{4\rho^2} \Big( (J_1+J_2)+(J_1-J_2) \cos\theta\Big) d\psi + \dots
}
where $\theta$ is the angle between $\vec r$ and the dipoles $\vec{D} \equiv \sum_{j=1}^N \sum_{I=1}^3 \kt^I_j \vec r_j$. The two angular momenta are then given by
\eqn{ 
J_R \equiv J_1+J_2 = \frac{4}{3} C_{IJK} \sum_{j=1}^N \frac{\kt^I_j \kt^J_j \kt^K_j}{v_j^2} \quad \text{and} \quad
J_L \equiv J_1-J_2 = 8 |\vec{D}|.
\label{eq:angmomenta}
}
Using the bubble equations, one can associate an angular momentum flux vector with the $ij^{\rm th}$ bubble:
\eqn{ 
\vec J_L = \sum_{i>j} \vec{J}_{L, ij}\,,\qquad 
\vec{J}_{L, ij} \equiv - \frac{4}{3} v_i v_j C_{IJK} \Pi^{(I)}_{ij} \Pi^{(J)}_{ij} \Pi^{(K)}_{ij} \frac{(\vec r_i-\vec r_j)}{|\vec r_i-\vec r_j|}.
}
The flux on the left-hand side of the bubble equation (\ref{eq:bubbleeq}) yields the contribution of the bubble to $J_L$.

\section{Non-extremal black hole geometry}\label{app:BlackHole}

The non-extremal rotating black hole solution sourced by three M2's on $T^6$ is the Cvetic-Youm black hole. We give it in the notation of \cite{Chowdhury:2011qu}. The solution depends on six parameters: a mass parameter $m$, three `boosts' $\delta_I$ related to the charges and angular momentum parameters $a_1,a_2$. The metric is
\be
ds_{11}^2 = -(H_1 H_2 H_3)^{-2/3} H_m (dt + k)^2 + (H_1 H_2 H_3)^{1/3} ds_4^2 + \sum_{I=1}^3 \frac{(H_1 H_2 H_3)^{1/3}} {H_I}ds_I^2\,.\label{eq:BH_Metric}
\ee
with
\be
k = \f{m}{f} \left[ -\frac{ c_1 c_2 c_3}{H_m} (a_1 \cos^2\theta\, d\psi + a_2\sin^2 \theta\, d\phi) + s_1s_2s_3 (a_2 \cos^2\theta\, d\psi + a_1\sin^2 \theta \,d\phi) \right]
\ee
with $I,J,K$ all different and we write
\be
c_I \equiv \cosh \delta_I\,, \qquad s_I \equiv \sinh \delta_I\,.
\ee
The solution is built from the functions
\be
H_I = 1 + \frac{m s_I^2}{f}\,,\qquad H_m = 1 - \frac{m}f\,, \qquad f = \rho^2 + a_1^2 \sin^2\theta + a_2^2 \cos^2 \theta\,.
\ee
The four-dimensional metric is 
\bea
ds_4^2 &=& \frac{f \rho^2}{g}d\rho^2 + f ( d \theta^2 + \sin^2\theta\, d \phi^2 + \cos^2 \theta\, d\psi^2)\nn
&&+ H_m^{-1} (a_1 \cos^2\theta \,d\psi + a_2\sin^2 \theta d\phi)^2 - (a_2 \cos^2\theta \,d\psi + a_1\sin^2 \theta \,d\phi)^2\,,\nn
g &=& (\rho^2 + a_1^2)(\rho^2 + a_2^2) - m \rho^2 \equiv (\rho^2 - \rho^2_+)(\rho^2 - \rho^2_-)\,.\label{eq:4d_Base}
\eea
The inner and outer horizon are given by the roots of $g(\rho)$:
\be
(\rho_\pm)^2= \frac 12 \left( m- {a_1^2}-{a_2^2} \pm \sqrt{\left(m- a_1^2-a_2^2\right)^2-4 a_1^2 a_2^2}\right)\,.
\ee
The ADM mass, electric charges and angular momenta of the black hole are 
\eal{
M_{ADM} &= \frac m 2 \sum_I \cosh 2 \delta_I\,, \qquad &J_1 =m( a_1 c_1 c_2 c_3 - a_2 s_1 s_2 s_3)\,,\\
\bQ_I &= \frac m 2 \sinh 2 \delta_I\,, &J_2 =-m( a_2 c_1 c_2 c_3 - a_1 s_1 s_2 s_3)\,,
\label{eq:5dCharges_app}
}
where we have set $G_5 = \f{\pi}{4}$ as discussed in appendix A of  \cite{Chowdhury:2011qu}.

\section{Approximation for  throat depth}\label{app:Depth}

We can give a good measure of the throat depth by integrating along the $z$-axis, from the outermost center $z_{MS} \equiv z_7$ up to a suitable cutoff scale $z_{\rm neck}$:
\be
L_{MS} \equiv \int_{z_{MS}}^{z_{\rm neck}} V ^{1/2}(Z_1 Z_2 Z_3)^{1/6} dz\,,\label{eq:MS_depth}
\ee
The depth of the black hole throat is the metric distance from the horizon at $\rho = \rho_+$ to the end of the throat at $\rho = \rho_{\rm neck}$ which can be approximated by
\be
\rho_{\rm neck} = (Q^{BH}_1 Q^{BH}_2 Q^{BH}_3)^{1/6}\,.
\ee
The depth of the throat is then given by integrating $\sqrt{g_{\rho\rho}}$ in the metric \eqref{eq:BH_Metric}:
\be
L_{BH} \equiv \int_{\rho_+}^{\rho_{\rm neck}} \sqrt{g_{\rho\rho}} d\rho = \int_{\rho_+}^{\rho_{\rm neck}}\frac{\rho\sqrt{f} }{\sqrt{g}} (H_1 H_2 H_3)^{1/6} d\rho \,.\label{eq:BH_Depth}
\ee
To get a feeling for $\Delta L$, we make some approximations. First, we approximate the geometry of the non-extremal black hole by a non-rotating one, so $a_1 = a_2 = 0$. We get:
\be
L_{BH} = \int^{r_{\rm neck}}_{r_+} \frac{(H_1 H_2 H_3)^{1/6}}{\sqrt{1 - \frac m {\rho^2}}} d\rho\,,
\label{eq:Lbh}
\ee
with $H_I = 1 + Q^{BH}_I/\rho^2$. For near-extremal black holes, this is a good approximation. We also replace the microstate geometry by the (spherically symmetric) geometry of the extremal black hole metric: 
\be
L_{MS}=\int^{r_{\rm neck}}_{r_{MS}}{(Z_1 Z_2 Z_3)^{1/6} }\frac{d\rho}{\sqrt{r}} = \int^{\rho_{\rm neck}}_{\rho_{MS}}(Z_1 Z_2 Z_3)^{1/6} d\rho\,,
\label{eq:Lms}
\ee
where we performed the change of variables $r = \tfrac 14 \rho^2$ and we have $Z_I = 1 + Q_I/\rho^2$.
This is a valid approximation, since the extremal black hole geometry only differs significantly from the microstate very deep down the throat.

Second, we approximate the black hole integral by splitting it into a part where $\rho_{\rm neck} >\rho\gg \rho_+$ and a part where $\rho_{\rm neck} \gg \rho > \rho_+$. We choose some intermediate radius $\rho_{\rm int} \approx \sqrt{\rho_+ \rho_{\rm neck}}$, but its exact value is of no importance.\footnote{In our example, we have $\rho_{\rm neck}^2 \approx 10^6, \rho_+^2 \lesssim 10^2$ and we can choose $\rho_{\rm int}^2\sim 10^4$.} With this approximation \ref{eq:Lbh} becomes:
\bea
L_{BH}=\rho_{\rm neck}\int^{\rho_{\rm int}}_{\rho_+} \frac{ d\rho}{\sqrt{\rho^2 - \rho_+^2}}+\int^{\rho_{\rm neck}}_{\rho_{\rm int}} (H_1 H_2 H_3)^{1/6} d\rho \,,
\eea
where we used that for the non-rotating non-extremal black hole the non-extremality parameter $m$ is just the square of the horizon radius $\rho_+$. In the same way we approximate \ref{eq:Lms}:
\bea
L_{MS}= \rho_{\rm neck}\int^{\rho_{\rm int}}_{\rho_{MS}} \frac{ d\rho}{\rho}+ \int^{\rho_{\rm neck}}_{\rho_{\rm int}} (Z_1 Z_2 Z_3)^{1/6} d\rho\,.
\eea

Third, we know that  for the extremal and non-extremal black hole the charges are almost equal because we are working with supertube \textit{probes}, and hence also the $Z_I = H_I$ are equal. Then the difference in depth is
\bea
L_{BH} - L_{MS} &=&\rho_{\rm neck}\left(\int^{\rho_{\rm int}}_{\rho_+} \frac{ d\rho}{\sqrt{\rho^2 - \rho_+^2}}- \int^{\rho_{\rm int}}_{\rho_+}  \frac{d\rho}{r}\right)\nonumber\\
&=&\rho_{\rm neck} \left[ \ln \left(\frac{\rho_{\rm int} + \sqrt{\rho^2_{\rm int} - \rho_+^2}}{\rho_+}\right) - \ln \frac{\rho_{\rm int}}{\rho_{MS}}\right]
\eea
Since we chose $\rho_{\rm int} \gg \rho_+$, we can approximate this very well by
\be
\Delta L = L_{BH} - L_{MS} = \rho_{\rm neck} \ln \left(2\frac{\rho_{MS}}{\rho_{+}}\right)\,.
\ee

We can use this result to get some idea on how the size of the supertubes and the depth of the microstates affect $\Delta L$. Consider the scaling of the supertube charges and the coordinates of the microstate centers as
\bea
(\pQ_1,\pQ_2,d_3) &\to& e^\lambda (\pQ_1,\pQ_2, d_3)\,,\nonumber\\
\rho_{MS} &\to& e^{\mu} \rho_{MS}\,,\label{eq:scalings2}
\eea                                                                                                 
Both scalings have a non-trivial effect on the size of the horizon radius of the would-be non-extremal black hole, which (neglecting rotation) is given by
\be
\rho_+^2 = m = \sqrt{\frac{8\Delta M}{\frac 1{Q_1} + \frac 1{Q_2} + \frac 1{Q_2}}}\,.\label{eq:HorizonRadius}
\ee
The potential $\calh$ scales linearly with the tube charges, and in the scaling regime it also scales linearly with the coordinate size of the centers, see eq.\ \eqref{eq:ScalingHamiltonian}. The same applies, of course, to the value $\Delta M$ of its metastable minimum. Hence  the horizon radius, as a function of the tube charges $\pQ^{\rm tube} \equiv (\pQ_1,\pQ_2,d_3)$ and the size of the microstate background $\rho_{MS}$, has the following scaling behavior: 
\be
\rho_+(\pQ^{\rm tube};\rho_{MS}) = e^{-(\lambda + \mu)/4}\rho_+ (e^\lambda \pQ^{\rm tube};e^\mu \rho_{MS})  \,.
\ee
Therefore under the scalings \eqref{eq:scalings2}, the difference in depths $\Delta L \equiv L_{BH} - L_{MS}$ goes as
\be
\frac{\Delta L}{\rho_{\rm neck}} \to \frac{\Delta L}{\rho_{\rm neck}}- \frac{1}4\lambda+  \frac 3 4 \mu\,.\label{eq:ThroatScaling2}
\ee

}

\bibliographystyle{toine}
\bibliography{Supertube.bib}

\end{document}